\documentclass[aps,prl,amsmath,floats,floatfix,twocolumn,
  superscriptaddress,nofootinbib,showpacs]{revtex4-1}
\pdfoutput=1 
\usepackage{amsmath}    
\usepackage{lipsum}
\usepackage{amssymb}
\usepackage{amsfonts}
\usepackage{amsthm}
\usepackage{braket}
\usepackage{bm}
\usepackage{multirow}
\usepackage{color} 
\usepackage{array}
\usepackage{caption}

\usepackage{enumitem}
\usepackage{dcolumn}
\usepackage[dvipsnames]{xcolor}
\usepackage{epsfig}
\usepackage{graphicx}   
\usepackage{graphics}
\usepackage[latin1]{inputenc}
\usepackage{latexsym}
\usepackage{mathrsfs}
\usepackage{mathtools}
\usepackage{orcidlink}
\usepackage{rotating}
\usepackage{setspace}
\usepackage{siunitx}
\usepackage{amsmath,amssymb}
\usepackage{tikz}
\usetikzlibrary{arrows.meta,calc,positioning}
\usepackage{subfigure}  
\usepackage{url}
\usepackage{verbatim}	
\usepackage{hyperref}   

\hypersetup{
    colorlinks=true,
    urlcolor=blue,
    citecolor=blue,
    linkcolor=red
}
\definecolor{IsoGreen}{RGB}{0,120,70}
\definecolor{IsoRed}{RGB}{178,34,34}
\definecolor{HeaderGray}{RGB}{242,242,242}
\definecolor{SoftGreen}{RGB}{246,252,248}
\definecolor{SoftRed}{RGB}{255,247,247}

\usepackage{amsmath,amssymb,bm,mathtools}
\usepackage[T1]{fontenc}
\usepackage{lmodern}

\usepackage{amsthm}

\usepackage{xspace} 
\usepackage{cleveref}
\usepackage[toc,page]{appendix}
\usepackage{array}

\usepackage{yfonts}
\usepackage{tikz}
\usetikzlibrary{shapes.geometric, arrows}
\tikzstyle{null} = [rectangle, 
minimum width=0.5cm, 
minimum height=0.5cm, 
text centered, 
fill=white!30]
\tikzstyle{nullblue} = [rectangle, 
minimum width=0.5cm, 
minimum height=0.5cm, 
text centered, 
text = blue,
fill=white!30]
\tikzstyle{nullor} = [rectangle, 
minimum width=0.5cm, 
minimum height=0.5cm, 
text centered, 
text = orange,
fill=white!30]
\tikzstyle{nullred} = [rectangle, 
minimum width=0.5cm, 
minimum height=0.5cm, 
text centered, 
text = red,
fill=white!30]
\tikzstyle{nullsmall} = [rectangle, 
minimum width=0.1cm, 
minimum height=0.1cm]
\tikzstyle{bracket} = [rectangle, 
minimum width=0.3cm, 
minimum height=3cm]
\tikzstyle{ColinStyle} = [rectangle, 
minimum width=1.2cm, 
minimum height=1cm, 
text centered, 
draw=black, 
line width = 0.7pt,
fill=white!30]
\tikzstyle{arrow} = [line width = 1.3 pt,->,>=stealth]
\tikzstyle{line} = [line width = 0.9pt,-,>=stealth]
\tikzstyle{bluethickline} = [line width = 2pt,-,>=stealth, color = blue]
\tikzstyle{orthickline} = [line width = 2pt,-,>=stealth, color = orange]
\tikzstyle{redthickline} = [line width = 2pt,-,>=stealth, color = red]
\makeatletter
\newcommand*{\rom}[1]{\expandafter\@slowromancap\romannumeral #1@}

\makeatother
\allowdisplaybreaks

\newcommand{\CAL}{Theoretical Astrophysics 350-17, California Institute of Technology, Pasadena, CA 91125, USA}
\newcommand{\ILL}{Illinois Center for Advanced Studies of the Universe \& Department of Physics,
University of Illinois at Urbana-Champaign, Urbana, Illinois 61801, USA}

\begin{document}
    \title{Parity Isospectrality of the Kerr--Newman--de Sitter Black Hole}
    \author{Colin Weller\,\orcidlink{0000-0001-5173-5638}}
    \email{cweller@caltech.edu}
    \affiliation{\CAL}
    
    \author{Dongjun Li\,\orcidlink{0000-0002-1962-680X}}
    \affiliation{\ILL}

    \author{Pratik Wagle\,\orcidlink{}}\affiliation{\ILL}
    
    \author{Yanbei Chen\,\orcidlink{0000-0002-9730-9463}}
    \affiliation{\CAL}
    
    \author{Nicol\'{a}s Yunes\,\orcidlink{} }
    \affiliation{\ILL}
    
    \date{\today}
    \begin{abstract}
    Vacuum Kerr perturbations in general relativity have identical even- and odd-parity quasinormal frequencies. We establish this parity isospectrality for linear gravitoelectromagnetic perturbations of subextremal Kerr--Newman--de Sitter black holes, without a slow-rotation or small-charge expansion. We reconstruct the radiative metric and gauge-potential perturbations directly from gauge-invariant master variables in radiation gauge, without an intermediate complex potential.
    
    \end{abstract}
    \maketitle
        

\noindent\textit{Introduction}---In four-dimensional Einstein--Maxwell theory, stationary, asymptotically
flat black holes in the electrovacuum belong to the
Kerr--Newman family under the standard uniqueness assumptions
\cite{Newman:1965my,Mazur:1982db,bunting1983proof}.
For vanishing magnetic charge, this family is characterized solely by the mass $M$,
angular momentum $J$, and electric charge $Q$. Its counterpart with a positive
cosmological constant $\Lambda_c$\footnote{The cosmological constant will be denoted $\Lambda_c$, while the NP Ricci scalar will be denoted $\Lambda_{\rm NP}$.} is the Kerr--Newman--de~Sitter solution
\cite{carter1968new}. Charged black holes provide a natural setting for studying
strong-field geometrodynamics \cite{Wheeler:1962kl,Nichols:2011pu, Zhang:2012jj,Nichols:2012jn, Boyeneni:2025tsx}, and the excitation of quasinormal modes, where
gravitational and electromagnetic perturbations are dynamically
coupled at linear order both in the nonrotating \cite{Zerilli:1974ai,de1987gravitational, Chandrasekhar_1983} and rotating \cite{lee1976coupled, torres1987complex, Pani:2013ija, Dias:2015wqa} cases. 

Black-hole perturbation theory was first developed for Schwarzschild
black holes in the Regge--Wheeler--Zerilli formalism
\cite{Regge:1957td,Zerilli:1970wzz}, where spherical symmetry reduces
the equations for even- and odd-parity perturbations to one-dimensional
Schr\"{o}dinger-type equations. Despite their different effective
potentials, the two sectors share the same quasinormal-mode spectrum---a property
known as \textit{parity isospectrality}
\cite{Chandrasekhar:1975zza,Chandrasekhar:1979iz}.
This equivalence was established through a Darboux-type transformation
of the equations of motion and boundary conditions
\cite{Chandrasekhar_1983,Anderson:1991kx,Glampedakis:2017rar},
a method subsequently extended to Reissner--Nordstr\"{o}m
\cite{Zerilli:1974ai,Chandrasekhar:1979iz},
Schwarzschild--de~Sitter \cite{GuvendeSitter}, and
Reissner--Nordstr\"{o}m--de~Sitter \cite{Mellor:1989ac} black holes.

However, for rotating backgrounds linear perturbations are usually formulated in the spinor calculus of Newman and Penrose
\cite{Newman:1961qr}. For the Kerr black hole, gravitational radiation in general relativity is described through the Weyl scalars $\Psi_0$ and $\Psi_4$, obeying the decoupled Teukolsky equations
\cite{Teukolsky:1972my, Teukolsky:1973ha, Press:1973zz, Teukolsky:1974yv}, which also generalize to perturbations of Kerr--de Sitter \cite{Khanal:1983kk,Suzuki:1998vy,Chambers:1994ap,Berens:2025kkm}. Remarkably, linear perturbations of Kerr are also isospectral \cite{Li:2023ulk}, which was shown by using
the parity-conjugation relations of the Teukolsky equations
\cite{Chrzanowski:1975wv,Casals:2005kr,Nichols:2012jn,Miguel:2023rzp}, and, additionally, radiation-gauge metric reconstruction from the Weyl scalars
\cite{Chrzanowski:1975wv,Chrzanowski:1976jy,Cohen_Kegeles_1975,
Kegeles:1979an,Wald:1978vm,Whiting:2005hr}. Some studies have connected black-hole isospectrality to an underlying
complex structure at the light ring, but only in the eikonal limit \cite{Pereniguez:2026avs,Bah:2026aia}. More recently,  Ref.~\cite{Weller:2026jqu} developed a unified
framework for parity isospectrality based on the complex linearity of the perturbation equations and boundary conditions, extending previous results to Kerr--de Sitter black holes and to all mode numbers, without relying on the eikonal limit. 

Using the Newman--Penrose formalism, quasinormal modes of Kerr--Newman black holes have been studied using 
the Dudley--Finley approximation
\cite{Dudley:1977zz,dudley1979covariant,Kokkotas:1993ef,Berti:2005eb,Saha:2025nsg}. Other studies have  retained the gravitoelectromagnetic coupling, while making a slow-rotation
\cite{Pani:2013wsa,Pani:2013ija} or small-charge-to-mass-ratio
\cite{Mark:2014aja} approximation. At generic spin and charge, the fully coupled Newman--Penrose equations
can be expressed in terms of two gauge-invariant master variables
\cite{torres1987complex,Dias:2015wqa}\footnote{Chandrasekhar derived a
related system in a fixed tetrad gauge, known as the phantom gauge\cite{Chandrasekhar_1983}.}.
Although numerical evidence supports parity isospectrality at first
\cite{Pani:2013ija,Pani:2013wsa} and second
\cite{Blazquez-Salcedo:2022eik} orders in spin, demonstrating that 
isospectrality persists for generic subextremal spins has remained an open problem.

In this Letter, we establish that 
gravitoelectromagnetic perturbations of subextremal Kerr--Newman--de~Sitter black holes are parity isospectral,
without a slow-rotation or small charge expansion. We first extend the coupled
master system for Kerr--Newman perturbations
\cite{lee1976coupled,Dias:2015wqa} to the Kerr--Newman--de~Sitter background.
We then reconstruct both the metric and gauge-potential perturbations
from its solutions in the ingoing radiation gauge by directly integrating
the linearized Newman--Penrose identities, without introducing an intermediate complex potential. We use this reconstruction to show that the theorem of Ref.~\cite{Weller:2026jqu} applies,
from which parity isospectrality follows.

\vspace{0.2cm}
\noindent\textit{Formalism}---We work in standard Einstein--Maxwell theory with fixed cosmological constant $\Lambda_c$, signature $(-,+,+,+)$ and geometric units where $G = c =1 $. The Einstein--Maxwell action is given by
\begin{align}\label{eq: EinsteinMaxwellAction}
    S &= \kappa_g \int \,d^4x \sqrt{|g|}\left[R - 2 \Lambda _c - F_{\mu\nu}F^{\mu \nu}\right]\,,
\end{align}
where $\kappa_g = (16\pi)^{-1}$, $R$ is the Ricci scalar, $\Lambda_c$ is the cosmological constant, $F_{\mu \nu} = 2\nabla_{[\mu}A_{\nu]}$ is the Maxwell field-strength tensor, where $A_{\mu}$ is the Maxwell gauge field and the brackets $[\mu,\nu]$ denote the standard antisymmetrization. Henceforth, Greek and Latin letters used as tensor indices label spacetime components and components in the null tetrad respectively;
unindexed symbols, such as $\kappa$, $\sigma$, $\rho$, $\mu$, and $\nu$,
denote Newman--Penrose spin coefficients.

We take the background spacetime to be the Kerr--Newman--de~Sitter solution of the Einstein--Maxwell equations [see Eq.~\eqref{eq: EinsteinMaxwellEquOfMot}] with a positive cosmological constant $\Lambda_{\rm c}$. In Boyer--Lindquist coordinates $(t,r,\theta,\varphi)$ \cite{Boyer:1966qh}, the line element of the spacetime and the corresponding gauge potential are given in Eq.~\eqref{eq: KNdSLineElemenet} and Eq.~\eqref{eq: KNdSGaugeField} of the End Matter, respectively. This spacetime is classified as Petrov-type D \cite{Petrov:2000bs}, so we can fix a null tetrad [see Eq.~\eqref{eq: KNDSFrame}] $\{l^{\mu},n^{\mu},m^{\mu},\bar{m}^{\mu}\}$, which is compatible with the spacetime, i.e., $g_{\mu \nu} = - 2l_{(\mu}n_{\nu)} + 2m_{(\mu}\bar{m}_{\nu)}$, with $l^{\mu}$ and $n^{\mu}$ aligned with the two repeated outgoing and ingoign principal null directions, respectively. In this spacetime, we have the following results \cite{Chandrasekhar_1983}:
\begin{align}
\Psi_0=\Psi_1=\Psi_3=\Psi_4=0\,, &\quad \kappa=\sigma=\nu=\lambda=0\,,\\
    \Phi_0 = &\Phi_2 = 0\,,\nonumber
\end{align}
where $\Psi_i$ are self-dual Weyl scalars and $\Phi_{0,2}$ are extreme-spin projections of the Maxwell field-strength tensor\footnote{Note that $\Phi_{0,1,2}$ denotes Maxwell scalars while $\Phi_{ij}$, where $i,j \in \{0,1,2\}$ are Newman--Penrose Ricci scalars.} [see Eq.~\eqref{eq: MaxwellNPScalars}]. 
In this frame, the only non-vanishing Weyl scalar is the Coulomb-type Weyl scalar $\Psi_2$ and Maxwell scalar $\Phi_1$ [see Eq.~\eqref{eq:KNdSBackgroundScalars}].

We consider simultaneous, real-valued, linear metric and gauge potential perturbations about the Kerr--Newman--de~Sitter background of the form $g_{\mu \nu} + \epsilon h^{(1)}_{\mu \nu}$ and $A_{\mu }+ \epsilon A^{(1)}_{\mu}$, where $\epsilon \ll 1$ is a real-valued bookkeeping parameter. We will use a superscript $(1)$ to denote perturbed quantities that enter equations at $\mathcal{O}(\epsilon^1)$, while quantities with no superscript are assumed to contribute at the background, $\mathcal{O}(\epsilon^0)$, level. The metric $h^{(1)}_{\mu \nu}$ and gauge potential $A^{(1)}_{\mu}$ perturbations must jointly satisfy the linearized equations of motion [see \eqref{eq: EinsteinMaxwellEquOfMot}]:
\begin{subequations}\label{eq: LinearEquOfMot}
\begin{align}
        G^{(1)}_{\mu \nu}[h] + \Lambda_c h^{(1)}_{\mu \nu} &= 8 \pi T^{(1)}_{\mu \nu}\\
    \left(\nabla^{\mu} F_{\mu \nu}\right)^{(1)} &= 0
\end{align}
\end{subequations}
where $G^{(1)}_{\mu \nu}[h]$ is the linearized Einstein tensor, obtained by varying $G_{\mu \nu} \equiv R_{\mu \nu} - Rg_{\mu \nu}/2$ about the Kerr--Newman--de~Sitter solution, $T^{(1)}_{\mu \nu}$ is the perturbed Maxwell stress-energy tensor and $(\nabla^{\mu}F_{\mu \nu})^{(1)}$ is the linearized Maxwell current. 

Similar to the derivation of the Teukolsky equation, Eq.~\eqref{eq: LinearEquOfMot} can be projected onto the Newman--Penrose
basis to obtain coupled equations in terms of gauge-invariant curvature
variables. For Kerr--Newman, Refs.~\cite{lee1976coupled,Dias:2015wqa} demonstrated the radiative perturbations are governed
by a closed system of two coupled second-order partial differential
equations. Here we extend this construction to a nonvanishing
cosmological constant. Evaluating the resulting equations on the
Kerr--Newman--de~Sitter background yields (see Supplemental Material \cite{SupplementalMaterial} for the derivation)
\begin{subequations}\label{eq: OutgoingKNEqus}
\begin{align}
\left(\mathcal{O}_{+2}
+\Phi_{11}^{(0)}\mathcal{P}_{+2}\right)
\varphi^{(1)}_{+2}
+\Phi_{11}^{(0)}\mathcal{Q}_{+2}
\varphi^{(1)}_{+1}
&=0\,,
\\
\left(\mathcal{O}_{+1}
+\Phi_{11}^{(0)}\mathcal{P}_{+1}\right)
\varphi^{(1)}_{+1}
+\Phi_{11}^{(0)}\mathcal{Q}_{+1}
\varphi^{(1)}_{+2}
&=0,
\end{align}
\end{subequations}
where the corresponding gauge- and tetrad-invariant master variables are
\begin{align}\label{eq: OutgoingMasterKN}
\varphi^{(1)}_{+2}
&=\Psi^{(1)}_{0},
&
\varphi^{(1)}_{+1}
&=2\Phi^{(0)}_{1}\Psi^{(1)}_{1}
 -3\Psi^{(0)}_{2}\Phi^{(1)}_{0},
\end{align}
and the second-order differential operators
$\mathcal{O}_{+s}$, $\mathcal{P}_{+s}$, and
$\mathcal{Q}_{+s}$ are given in the End Matter
[see Eq.~\eqref{eq: PosistiveSpinOperators}].
For $\Lambda_c\to0$, Eq.~\eqref{eq: OutgoingKNEqus}
reduces to the Kerr--Newman system of
Ref.~\cite{Dias:2015wqa}.
In the uncharged limit
$Q\to0$, for which $\Phi_{11}^{(0)}\to0$, the
gravitoelectromagnetic coupling vanishes and the two equations reduce
to the spin-$2$ and spin-$1$ Teukolsky equations on
Kerr--de~Sitter \cite{Khanal:1983kk,Suzuki:1998vy};
taking in addition $\Lambda_c\to0$ recovers the standard Kerr
Teukolsky equations \cite{Teukolsky:1972my,Teukolsky:1973ha}.

\vspace{0.2cm}
\noindent \textit{Reconstruction of gravitoelectromagnetic perturbations}---Our goal will be to reconstruct the metric perturbation $h^{(1)}_{\mu \nu}$ and the gauge potential $A^{(1)}_{\mu}$ from the master variables $(\varphi^{(1)}_{+2},\varphi^{(1)}_{+1})$ satisfying Eq.~\eqref{eq: OutgoingKNEqus}. We denote metric perturbations in the Newman--Penrose basis by $h^{(1)}_{ab} = h^{(1)}_{\mu \nu}e^\mu_ae^\nu_b$ and gauge potential perturbations by $A^{(1)}_{a} = A^{(1)}_{\mu}e^\mu_a$, where $e^{a}_{\mu}$ is in the set $\{l^{\mu},n^{\mu},m^{\mu},\bar{m}^{\mu}\}$. For example $h^{(1)}_{nm} = h^{(1)}_{\mu \nu}n^{\mu}m^{\nu}$ and $A^{(1)}_{m} = A^{(1)}_{\mu}m^{\mu}$. In this basis, there are four real  $\{h^{(1)}_{nn},h^{(1)}_{ll},h^{(1)}_{ln},h^{(1)}_{m\bar{m}}\}$ and three complex $\{h^{(1)}_{nm},h^{(1)}_{mm},h^{(1)}_{lm}\}$ independent components of the metric perturbation along with one complex $\{A^{(1)}_{m}\}$ and two real $\{A^{(1)}_{n},A_{l}\}$ components in the gauge potential. Since we are considering real-valued solutions the perturbations satisfy $\overline{A^{(1)}_{m}} = A^{(1)}_{\bar{m}}$ and $\overline{h^{(1)}_{mm}} = h^{(1)}_{\bar{m}\bar{m}}$, where an overbar denotes complex conjugation.

First, we establish that linear perturbations of the Kerr--Newman--de~Sitter black hole in Einstein-Maxwell gravity  can be expressed in a traceless radiation gauge. The ingoing radiation gauge satisfies
\begin{equation}
    h_{\mu\nu}^{(1)} l^\mu = 0\,,
    \qquad
    g^{\mu\nu} h_{\mu\nu}^{(1)} = 0\,.
\end{equation}
Replacing the outgoing null four-vector  $l^\mu$ by the ingoing null four-vector $n^{\nu}$ defines the outgoing radiation gauge. Hereon, our reconstruction procedure will be in the ingoing radiation gauge. 

Since the Einstein--Maxwell action is invariant under diffeomorphisms, we can use the four real coordinate-gauge degrees of freedom in the gauge four-vector $\xi_{\mu}$ to set $h^{(1)}_{\mu \nu}l^{\mu} = 0$ by solving $(h^{(1)}_{\mu \nu} - 2\nabla_{(\mu}\xi_{\nu)})l^{\mu} = 0$. Additionally, the linear stress-energy component $T^{(1)}_{\mu \nu} l^{\mu}l^{\nu}= T^{(1)}_{ll}$ vanishes, which allows the trace to be set to zero with residual gauge, exhausting the remaining coordinate-gauge freedom \cite{Price:2006ke}. Moreover, we can set $A^{(1)}_{l} = 0$  by solving $(A^{(1)}_{\mu} - \nabla_{\mu} g^{(1)})l^{\mu} = 0$, where $g^{(1)}$ is a scalar function of the spacetime coordinates.  Lastly, there remains six real degrees of freedom in the frame gauge  at $\mathcal{O}(\epsilon^1)$. For our reconstruction procedure, we use the following null frame \cite{Campanelli:1998jv, Loutrel:2020wbw}
\begin{subequations}\label{eq: FrameGaugeIRG}
    \begin{align}
        l^{\mu(1)}&=0\quad 
         n^{\mu(1)}=\frac{1}{2} h_{n n}^{(1)} l^\mu\,,\\ m^{\mu(1)}&=h_{n m}^{(1)} l^\mu-\frac{1}{2} h_{m m}^{(1)} \bar{m}^\mu\,,
    \end{align}
\end{subequations}
We make use of the following modified directional derivatives, first introduced in Ref.~\cite{Wagle:2023fwl}
\begin{subequations}
    \begin{align}
D_{[a, b, c, d]} & =D+a \varepsilon+b \bar{\varepsilon}+c \rho+d \bar{\rho}\,, \\
\Delta_{[a, b, c, d]} & =\Delta+a \mu+b \bar{\mu}+c \gamma+d \bar{\gamma}\,, \\
\delta_{[a, b, c, d]} & =\delta+a \bar{\alpha}+b \beta+c \bar{\pi}+d \tau\,, \\
\bar{\delta}_{[a, b, c, d]} & =\bar{\delta}+a \alpha+b \bar{\beta}+c \pi+d \bar{\tau}\,,
    \end{align}
\end{subequations}
where $\{D,\Delta,\delta,\bar{\delta}\}$ are the usual Newman--Penrose directional derivatives and $\{a,b,c,d\}$ are constants. The Newman--Penrose spin coefficient $\varepsilon$ is distinct from the perturbation bookkeeping parameter $\epsilon$.

We begin by considering the Ricci identity
\begin{equation}\label{eq:SigmaRicciIdentity}
\Psi_0=D_{[-3,1,-1,-1]} \sigma-\delta_{[-1,-3,1,-1]} \kappa\,,
\end{equation}
where $\sigma$ and $\kappa$ are the standard Newman--Penrose optical scalars. Expanding this identity to $\mathcal{O}(\epsilon^1)$ in the frame gauge of Eq.~\eqref{eq: FrameGaugeIRG} yields 
\begin{align}\label{eq: Psi0ToSigma}
\varphi_{+2}^{(1)}=D_{[-3,1,-1,-1]} \sigma^{(1)}\,.
\end{align}
Thus, with a solution $\varphi^{(1)}_{+2}$ from Eq.~\eqref{eq: OutgoingKNEqus}, then Eq.~\eqref{eq: Psi0ToSigma} is a first order transport equation for $\sigma^{(1)}$. We can relate $\sigma^{(1)}$ to the metric perturbation \cite{Li:2022pcy,Loutrel:2020wbw}
\begin{align}\label{eq: PerturbedSigmaIRG}
    \sigma^{(1)}=-\frac{1}{2} D_{[-2,2,1,-1]} h_{m m}^{(1)}\,,
\end{align}
Once $\sigma^{(1)}$ is known, Eq.~\eqref{eq: PerturbedSigmaIRG} is a linear first-order transport equation for $h^{(1)}_{mm}$ along the integral curves of $l^{\mu}$, since $D=l^{\mu}\nabla_{\mu}$. These equations can be solved by an integrating factor on each null curve. Specifying $h^{(1)}_{mm}$ on a hypersurface transverse to these curves fixes the homogeneous integration freedom and determines $h^{(1)}_{mm}$ locally.

The key next step is to consider the Bianchi identity of Eq.~\eqref{eq: Bianchihnm}, which after perturbations becomes
\begin{align}\label{eq: PerturbedhnmBianchi}
\bar{\delta}_{[-4,0,1,0]}\Psi^{(1)}_0 -D_{[-2,0,-4,0]}\Psi^{(1)}_1 &= -D_{[-2,0,0,-2]}\Phi^{(1)}_{01}\,,
\end{align}
where $\Phi^{(1)}_{01}$ is a perturbed Ricci scalar, which generically satisfies
\begin{align}\label{eq: PerturbedRicci}
\Phi^{(1)}_{ij} = 2\left(\Phi_i^{(1)}\bar{\Phi}_j + \Phi_i\bar{\Phi}_j^{(1)}\right)\,,
\end{align}
and $i,j$ are integers in the set $\{0,1,2\}$.
Using the background Newman--Penrose equations, $D\Phi_1 = 2 \rho \Phi_1$ and $D \Psi_2=3 \rho \Psi_2+2 \rho \Phi_{11}$, Eq.~\eqref{eq: PerturbedhnmBianchi} can be written in terms of $\varphi^{(1)}_{+1}$
\begin{align}\label{eq: Phi0Transport}
    D_{[0,0,-3,0]} \Phi_0^{(1)}=\frac{2 \Phi_1\bar{\delta}_{[-4,0,1,0]}\Psi_0^{(1)}-D_{[0,0,-6,0]} \varphi^{(1)}_{+1}}{3 \Psi_2-2 \Phi_{11}}\,,
\end{align}
from which $\Phi^{(1)}_0$ can be obtained because $\Psi_0^{(1)}$ is known through Eq.~\eqref{eq: OutgoingMasterKN}, while $\varphi^{(1)}_{+1}$ and $\varphi^{(1)}_{+2}$ have been solved for already.

Next, consider the following Ricci identity:
\begin{equation}\label{eq:RhoRicciIdentity}
\begin{split}
\delta_{[-1,-1,0,-1]} \rho-\bar{\delta}_{[-3,1,0,0]} \sigma+\tau \bar{\rho}&
\\{}-\kappa(\mu-\bar{\mu})&=-\Psi_1+\Phi_{01}\,,
\end{split}
\end{equation}
which to $\mathcal{O}(\epsilon^1)$ gives
\begin{align}
\delta^{(1)}_{[-1,-1,0,-1]} \rho - \bar{\delta}_{[-3,1,0,0]} \sigma^{(1)} +\tau^{(1)}\bar{\rho}=-\Psi^{(1)}_1+\Phi^{(1)}_{01}\,.
\end{align}
    Using the reconstructed quantities directly in terms of the metric components in the ingoing radiation gauge [see Eq.~\eqref{eq:ReconstructuedSpinCoeifEM} of the End Matter] leads to
    \begin{align}\label{eq: MasterhnmDiffEqu}
        \left[(2 \rho-\bar{\rho}) D+2 \rho^2-2 |\rho|^2+\bar{\rho}^2\right] h_{n m}^{(1)} = \mathcal{S}^{(1)}_{nm}[\sigma^{(1)}, \Psi^{(1)}_1,\Phi^{(1)}_{01}],
    \end{align}
    where $\mathcal{S}_{nm}^{(1)}[\sigma^{(1)}, \Psi^{(1)}_1,\Phi^{(1)}_{01}] \equiv 2(\bar{\delta}-3 \alpha+\bar{\beta}) \sigma^{(1)} -2 \Psi_1^{(1)}+2 \Phi^{(1)}_{01}$.  This source depends only on previously reconstructed quantities:
$\sigma^{(1)}$ is obtained from Eq.~\eqref{eq: Psi0ToSigma},
$\Psi_1^{(1)}$ is recovered from the definition of 
$\varphi_{+1}^{(1)}$ in Eq.~\eqref{eq: OutgoingMasterKN} (which requires $\Phi_0^{(1)}$), and
$\Phi_{01}^{(1)}$ is determined by Eqs.~\eqref{eq: Phi0Transport} and \eqref{eq: PerturbedRicci}.
Thus, Eq.~\eqref{eq: MasterhnmDiffEqu} reconstructs $h_{nm}^{(1)}$
through a first-order transport equation.

We now turn to reconstructing the complex-valued component $A^{(1)}_m$ of the gauge potential perturbation.
The Maxwell scalar $\Phi^{(1)}_0$ can be expressed explicitly in terms of components \cite{Wardell:2020naz}
\begin{align}\label{eq: Phi0TransportSolve}
\Phi^{(1)}_0
=
D_{[-1,1,0,-1]} A^{(1)}_m\,,
\end{align}
where we have used $A^{(1)}_{l} = 0$.
Therefore, with $\Phi^{(1)}_0$ obtained from Eq.~\eqref{eq: Phi0Transport}, $A^{(1)}_{m}$ is obtained by integrating the transport equation in Eq.~\eqref{eq: Phi0TransportSolve}.

Next, we compute $h^{(1)}_{nn}$.
Expanding one of the Maxwell equation [see Eq.~\eqref{eq:  MaxwellOneTwoEM} of the End Matter] to $\mathcal{O}(\epsilon^1)$ yields
\begin{equation}\label{eq:PhiOneMaxwellTransport}
(D-2 \rho)\Phi^{(1)}_1-\left(\bar{\delta}+\pi-2 \alpha\right)^{(0)} \Phi^{(1)}_0=0\,,
\end{equation}
which is a first order transport equation for $\Phi^{(1)}_1$ because $(D- 2\rho)^{(1)}$ vanishes in the ingoing radiation gauge. 
Consequently, from a solution of $\Phi^{(1)}_1$, we can now compute the Ricci sources $\Phi^{(1)}_{11}$ and $\Phi^{(1)}_{01}$. After linearizing the Bianchi identity [given in Eq.~\eqref{eq: Bianchihnn} of the End Matter], we obtain 
\begin{align}\label{eq: Psi2Transport}
    (D-3 \rho) \Psi^{(1)}_2 = \mathcal{S}^{(1)}_{\Psi_2}[\Psi^{(1)}_1,\Phi^{(1)}_{01},\Phi^{(1)}_{11},\Phi^{(1)}_{10}]\,,
\end{align}
where $\mathcal{S}^{(1)}_{\Psi_2}$ is given in Eq.~\eqref{eq: SourcePsi_2} of the End Matter. The solution of this first-order transport equation then yields $\Psi^{(1)}_2$. Furthermore, linearizing the Ricci identity given in Eq.~\eqref{eq: Riccihnn}
we obtain 
\begin{align}\label{eq: hnnTransport}
\rho D_{[0,0,1,-1]}h^{(1)}_{nn} = \mathcal{S}_{nn}[\Psi^{(1)}_2,h^{(1)}_{mm},h^{(1)}_{nm}]
\end{align}
where $\mathcal{S}_{nn}[\Psi^{(1)}_2,h^{(1)}_{mm},h^{(1)}_{nm}]$  is a source term that depends only the metric components $h^{(1)}_{nm}$ and $h^{(1)}_{mm}$, and on the Weyl scalar $\Psi^{(1)}_2$ [see Eq.~\eqref{eq: hnnSourceTerm}]. Solving this first order transport equation then yields $h_{nn}^{(1)}$. 
As a corollary, we can solve for $A^{(1)}_n$ using
\begin{align}\label{eq: AnTransport}
    \Phi^{(1)}_1=\frac{1}{2}\left[D_{[1,1,1,-1]} A^{(1)}_n+ \mathcal{S}^{(1)}_{A}\right]
\end{align}
where $S^{(1)}_{A} = \bar{\delta}_{[-1,1,-1,-1]} A^{(1)}_m - \delta_{[-1,1,1,1]} A^{(1)}_{\bar{m}}$. Since we have solved for $\{h^{(1)}_{mm},h^{(1)}_{nm},h^{(1)}_{nn},A^{(1)}_{m},A^{(1)}_{n}\}$, we have completed the reconstruction for the metric  $h^{(1)}_{\mu\nu }$ and gauge potential $A^{(1)}_{\mu}$ perturbations.

\usetikzlibrary{fit}

\begin{figure*}[t]
\centering
\resizebox{0.995\textwidth}{!}{%
\begin{tikzpicture}[
  x=1cm,y=1cm,font=\small,
  arrow/.style={
    -{Stealth[length=2.2mm,width=1.5mm]},
    line width=0.65pt
  },
  connector/.style={line width=0.65pt},
  field/.style={
    inner sep=1.5pt,
    minimum width=1.20cm,
    minimum height=0.45cm,
    align=center
  },
  eqslot/.style={
    fill=white,
    inner xsep=2pt,
    inner ysep=0.5pt,
    font=\scriptsize
  }
]

\node[
  draw,rectangle,
  inner xsep=8pt,inner ysep=5pt,
  align=center,font=\scriptsize
] (system) at (0.8,0.55)
  {Master variables\\[2pt]
   Eq.~\eqref{eq: OutgoingKNEqus}};

\coordinate (fork) at (2.25,0.55);
\draw[connector] (system.east)--(fork);
\fill (fork) circle[radius=0.9pt];

\node[field] (v2) at (3.4,1.50)
  {$\varphi_{+2}^{(1)}$};
\node[field] (v1) at (3.4,-0.40)
  {$\varphi_{+1}^{(1)}$};

\draw[arrow] (fork)|-(v2.west);
\draw[arrow] (fork)|-(v1.west);

\node[field] (sigma) at (5.7,1.50)
  {$\sigma^{(1)}$};
\node[field] (hmm) at (8.4,1.50)
  {$h_{mm}^{(1)}$};

\draw[arrow] (v2.east)--
  node[eqslot,above=3pt]
  {Eq.~\eqref{eq: Psi0ToSigma}}
  (sigma.west);

\draw[arrow] (sigma.east)--
  node[eqslot,above=3pt]
  {Eq.~\eqref{eq: PerturbedSigmaIRG}}
  (hmm.west);

\node[field] (Phi0) at (5.7,-0.40)
  {$\Phi_0^{(1)}$};
\node[field] (Ricci) at (8.4,-0.40)
  {$\Phi_{11}^{(1)},\ \Phi_{01}^{(1)}$};
\node[field] (Psi1) at (11.2,-0.40)
  {$\Psi_1^{(1)}$};
\node[field] (hnm) at (13.8,-0.40)
  {$h_{nm}^{(1)}$};

\draw[arrow] (v1.east)--
  node[eqslot,above=3pt] (eqPhi0Transport)
  {Eq.~\eqref{eq: Phi0Transport}}
  (Phi0.west);

\draw[arrow]
  (v2.south)--(3.4,0.50)
  -| (eqPhi0Transport.north);

\draw[arrow] (Phi0.east)--
  node[eqslot,above=3pt]
  {Eq.~\eqref{eq: PerturbedRicci}}
  (Ricci.west);
\draw[arrow] (Ricci.east)--
  node[eqslot,above=3pt]
  {Eq.~\eqref{eq: OutgoingMasterKN}}
  (Psi1.west);

\draw[arrow] (Psi1.east)--
  node[eqslot,above=3pt]
  {Eq.~\eqref{eq: MasterhnmDiffEqu}}
  (hnm.west);

\node[field] (Am) at (5.7,-2.50)
  {$A_m^{(1)}$};

\draw[arrow] (Phi0.south)--
  node[eqslot,right=3pt]
  {Eq.~\eqref{eq: Phi0TransportSolve}}
  (Am.north);

\node[field] (Psi2) at (11.2,-2.50)
  {$\Psi_2^{(1)}$};
\node[field] (hnn) at (13.8,-2.50)
  {$h_{nn}^{(1)}$};
\node[field] (An) at (16.4,-2.50)
  {$A_n^{(1)}$};

\draw[arrow] (Psi1.south)--
  node[eqslot,right=3pt]
  {Eq.~\eqref{eq: Psi2Transport}}
  (Psi2.north);

\draw[arrow] (Psi2.east)--
  node[eqslot,above=3pt]
  {Eq.~\eqref{eq: hnnTransport}}
  (hnn.west);

\draw[arrow] (hnn.east)--
  node[eqslot,above=3pt]
  {Eq.~\eqref{eq: AnTransport}}
  (An.west);

\end{tikzpicture}%
}
\caption{
Reconstruction of the metric and electromagnetic perturbations
from the coupled Kerr--Newman--de Sitter master equations in
ingoing radiation gauge. Equation~\eqref{eq: OutgoingKNEqus}
determines $\varphi_{+2}^{(1)}$ and $\varphi_{+1}^{(1)}$.
The arrows organize the reconstruction steps; each equation uses
all its previously determined inputs. 
}
\label{fig:KN-reconstruction-flowchart}
\end{figure*}
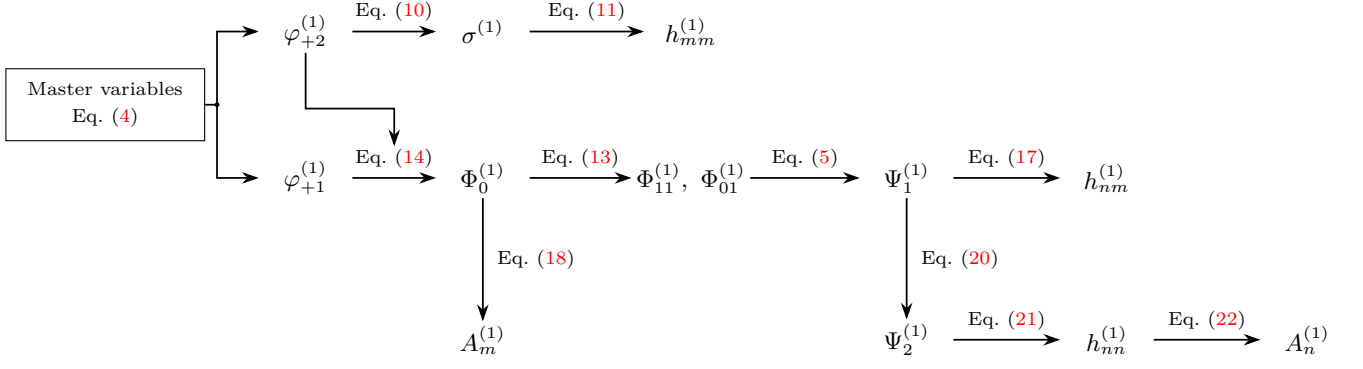

\vspace{0.2cm}
\noindent\textit{Isospectrality}---Having developed the reconstruction procedure, we now turn to demonstrating the gravitoelectromagnetic perturbations of Kerr--Newman--de~Sitter are parity isospectral. We apply 
Theorem~1B of Ref.~\cite{Weller:2026jqu}, which proves that the even- and
odd-parity quasinormal-mode frequencies coincide under the following conditions:
\begin{itemize}[nosep]
    \item[(i)] for every real-valued solution $(h^{(1)}_{\mu\nu},A^{(1)}_{\mu})$ there is a unique set of scalar master variables from which the radiative components are reconstructed, up to linear diffeomorphisms and a time translation; 
    \item[(ii)] both the equations of motion for the master variables and quasinormal-mode boundary conditions are
complex-linear; and 
    \item[(iii)] a complex-valued phase rotation $e^{i \chi}$, where $\chi$ is a constant in the open set $(0, \pi)$, of the master variables
maps every definite-parity solution to the opposite-parity.
\end{itemize}
When these criteria are satisfied, condition (i) ensures that the
reconstruction is complete and corresponds to a gauge-invariant physical
perturbation. Furthermore, condition (ii) guarantees that the phase rotation preserves the
quasinormal-mode frequency. Finally, condition (iii) ensures that this rotation maps each
physical quasinormal modes to one of the opposite parity. 

We first address condition (i). 
Two sets of reconstructed perturbations that differ only by a gauge
transformation lead to the same pair of gauge-invariant master variables.
To establish the converse, we must show that two pairs of perturbations
with identical master variables differ only by gauge. By linearity of the
reconstruction procedure and the master equations,
the difference between these reconstructed fields has vanishing master
variables, and thus, the master variables obey a homogeneous 
reconstruction hierarchy. The solution to the homogeneous first-order 
transport equations can then only depend on integration functions that 
are set by our choice of boundary conditions: purely ingoing at the 
future black-hole horizon and purely outgoing behavior at the future cosmological 
horizon. Imposing both horizon boundary conditions on the homogeneous
reconstruction prevents a QNM from being a solution since each step is a transport equation along an outgoing null direction. The two reconstructed 
metric--potential pairs therefore differ only by a gauge transformation. 
This establishes condition (i). 

Next, we show condition (ii) is satisfied. The equations of motion given in Eq.~\eqref{eq: OutgoingKNEqus} are complex-linear since neither equation couples the master variables to their complex conjugates $\bar{\varphi}^{(1)}_{+2}$ and $\bar{\varphi}^{(1)}_{+1}$. Additionally, quasinormal modes must satisfy Eq.~\eqref{eq: OutgoingKNEqus} together with
the correct boundary conditions, which for the Kerr--Newman--de~Sitter spacetime are also complex-linear, which we show next. Stationarity and axisymmetry allow the master variables to be decomposed in the Fourier domain as $\varphi^{(1)}_{s}
    =
    e^{-i\omega t+i m\phi}\,
    \mathcal{R}_{s}(r,\theta)\,,$
where $m$ is the azimuthal number and $\omega$ a complex-valued frequency.
Quasinormal modes are selected by imposing ingoing boundary conditions at the
future event horizon $r_+$, outgoing boundary conditions at the future
cosmological horizon $r_c$, and regularity of the eigenstates at the poles.  Near the two
horizons, the corresponding asymptotic behavior is given by
\begin{subequations}
\begin{align}
    \mathcal{R}_{s}
    &\sim
    (r-r_+)^{-s-i k_+/(2\kappa_+)}
    A^{H}_{s}(\theta),
    \label{eq:qnm_event}
    \\
    \mathcal{R}_{s}
    &\sim
    (r_c-r)^{-i k_c/(2\kappa_c)}
    A^{c}_{s}(\theta),
    \label{eq:qnm_cosmological}
\end{align}
\end{subequations}
where $s$ is the spin-weight, $\Omega_j=a/(r_j^2+a^2)$ is the angular velocity of each horizon, $\kappa_j=|\Delta_r'(r_j)|/[2\Xi(r_j^2+a^2)]$ is the magnitude
of the surface gravity, where the prime is a derivative with respect to $r$, and $k_j=\omega-m\Omega_j$ is the effective frequency in the
frame co-rotating with the corresponding horizon, where $j\in\{+,c\}$ labels the future horizon and the future cosmological horizon, respectively. 

The horizon conditions are complex-linear in the master variables since they are imposed on $\varphi^{(1)}_{+2}$ and $\varphi^{(1)}_{+1}$\footnote{In contrast, the quasinormal-mode boundary condition is not generically complex-linear for asymptotically anti-de Sitter spacetimes \cite{Dias:2013sdc,Weller:2026jqu}.}. 
Hence the quasinormal-mode boundary-value problem is complex-linear in the master variables.

Finally, let us address condition (iii). First, we use the metric reconstruction procedure to construct definite-parity metric--potential perturbations. Following Ref.~\cite{Li:2023ulk}, we
work in the canonical Boyer--Lindquist frame aligned with the black-hole
angular momentum, where the parity operator $\hat{P}$ acts on scalar functions as
$\hat{P}f(t,r,\theta,\phi)=f(t,r,\pi-\theta,\phi+\pi)$.
After angular-harmonic decomposition, definite-parity perturbations satisfy
$\hat{P}h^{(1)}_{\mu\nu}=\pm(-1)^\ell h^{(1)}_{\mu\nu}$ and
$\hat{P}A^{(1)}_\mu=\pm(-1)^\ell A^{(1)}_\mu$, where $\ell$ is the angular
momentum number and the upper (lower) sign denotes even (odd) parity
\cite{Regge:1957td,Zerilli:1970se,Moncrief:1974am,Nichols:2012jn,Li:2023ulk}. 

We also define complex conjugation by $\hat{C}f=\bar f$ and the
combined parity-conjugation operator by
$\hat{\mathcal{P}}=\hat{P}\hat{C}$.

Given definite-parity metric--potential perturbations, how do we construct definite-parity master functions? 
We begin by realizing that, using the parity-conjugation properties of the background Newman--Penrose quantities
(see the Supplemental Material), whenever
$(\varphi^{(1)}_{+2},\varphi^{(1)}_{+1})$ is a solution of Eq.~\eqref{eq: OutgoingKNEqus}, then so is
$(\hat{\mathcal{P}}\varphi^{(1)}_{+2},
-\hat{\mathcal{P}}\varphi^{(1)}_{+1})$. For a real, definite-parity solution
$(h_{\mu\nu}^{(1)},A_\mu^{(1)})$, the background tetrad transformation
$Pm^{\mu}=-\bar m^{\mu}$ implies
$\hat{\mathcal P}h_{mm}^{(1)}=\pm(-1)^\ell h_{mm}^{(1)}$
and
$\hat{\mathcal P}A_m^{(1)}=\mp(-1)^\ell A_m^{(1)}$. Since Eqs.~\eqref{eq: Psi0ToSigma} and \eqref{eq: PerturbedSigmaIRG} are even under a $\hat{\mathcal{P}}$ transformation, the corresponding definite parity mode for $\varphi^{(1)}_{+2}$ is given by $\varphi^{(1),\pm}_{+2}= \frac{1}{2}\left[
\varphi_{+2}^{(1)}
\pm(-1)^\ell\hat{\mathcal P}\varphi_{+2}^{(1)}
\right]$. Similarly, since Eqs.~\eqref{eq: Phi0Transport} and \eqref{eq: Phi0TransportSolve} are odd under a  $\hat{\mathcal{P}}$ transformation, the corresponding definite parity mode for $\varphi^{(1)}_{+1}$ is given by $\varphi^{(1),\pm}_{+1}= \frac{1}{2}\left[
\varphi_{+1}^{(1)}
\mp(-1)^\ell\hat{\mathcal P}\varphi_{+1}^{(1)}
\right]$. Importantly, the other reconstruction steps for the components $h_{nm}^{(1)}$, $h_{nn}^{(1)}$, and $A_n^{(1)}$
also respect the same $\hat{\mathcal P}$ transformation rules of $\varphi^{(1),\pm}_{+s}$. 

Now that we have definite-parity metric--potential solutions and master functions, we can establish condition (iii). Let us consider the chiral rotation of the master fields $e^{i \chi} = i$, so that we may write $\widetilde{\varphi}^{(1)}_{+s}=  e^{i \chi} \varphi^{(1)}_{+s} = i \varphi^{(1)}_{+s}$. These rotated master fields solve the same master equations 
and, through the reconstruction procedure, determine a new set of
real-valued metric--potential perturbations. Moreover, because the parity-conjugation operator $\hat{\mathcal P}$ is anti-linear, so that $\hat{\mathcal P}(i\varphi^{(1)}_{+s}) =-i\hat{\mathcal P}\varphi^{(1)}_{+s}$, the chirally-rotated fields have opposite parity relative to the un-rotated fields, i.e.~$\widetilde{\varphi}^{(1),\pm}_{+s} = i\varphi^{(1),\mp}_{+s}$. Therefore, by complex-linearity of Eq.~\eqref{eq: OutgoingKNEqus} and the boundary conditions, the two definite-parity eigenstates must have the same spectrum, and thus, condition (iii) is established.

\vspace{0.2cm}
\noindent\textit{Discussion}---In this Letter, we have shown the linear perturbations of the Kerr--Newman--de~Sitter black hole in Einstein--Maxwell theory are parity isospectral without an expansion in the spin, the charge, or the multipole of the radiation. In doing so, we have formulated a procedure that reconstructs the radiative components of the metric and gauge potential perturbations from the gauge invariant master functions $\varphi^{(1)}_{+2}$ and $\varphi^{(1)}_{+1}$. 

Beyond parity isospectrality, our reconstruction procedure extends
direct-integration methods to dynamically coupled degrees of freedom for the first time. This extension is nontrivial:
electromagnetic perturbations source the metric, while metric
perturbations enter Maxwell's equations, preventing independent
reconstruction of the two fields. Our closed reconstruction hierarchy
resolves this interdependence, determining both fields directly from
the gauge-invariant master variables.

Analogous to the Chrzanowski--Cohen--Kegeles approach
\cite{Chrzanowski:1975wv,Cohen_Kegeles_1975}, later formalized by Wald
\cite{Wald:1978vm}, the reconstruction of Ref.~\cite{torres1987complex}
expresses the metric and gauge potential perturbations in terms of
complex electromagnetic and gravitational potentials satisfying
a coupled system of equations. Using the parity-conjugate analysis on this method also yields the same definite-parity modes, and the same conclusion that each must be parity isospectral.  

Other methods that directly separate the perturbations for Kerr in general relativity 
\cite{Mei:2023pho,Mei:2026oem} rely on the conformally K\"ahler geometry of Kerr \cite{Aksteiner:2022bwr}. This suggests the possibility that one might be able to extend such arguments to the Kerr--Newman--de~Sitter background, thus avoiding metric reconstruction altogether. Recent progress has established the decoupling and separability of linear perturbations formulated in terms of curvature variables \cite{Hintz:2026jiw}, and similar structures may extend directly to metric perturbations.

 There are some notable advantages to our approach. In particular, we begin with gauge-invariant quantities, which are naturally connected to the observables entering a detector response. By contrast, the complex-potential formulation reaches these quantities only through an additional intermediate relation, making the connection to physical observables less direct. Additionally, it is not clear that generic prescribed sources at $\mathcal{O}(\epsilon^1)$ can be included in the procedures of Ref.~\cite{torres1987complex}, such as in Refs.~\cite{Green:2019nam, Toomani:2021jlo,Aksteiner:2016pjt,Bourg:2024vre,Casals:2024ynr}. In our procedure, however, sourced metric and gauge perturbations could be reconstructed using the traceful radiation gauge of 
Ref.~\cite{Li:2026rkf}.

Our techniques could also be used to study nonlinear
gravitoelectromagnetic ringdown, with the first-order
quantities providing the sources for the second-order Einstein--Maxwell
equations. Potential applications include studying how black-hole charge affects
the excitation of quadratic quasinormal modes \cite{Ma:2024qcv},
self-force calculations for extreme-mass-ratio inspirals
\cite{vandeMeent:2017bcc,ZimmermanPoisson2014},
and, in the asymptotically flat limit, the electromagnetic
contribution to gravitational-wave memory \cite{Christodoulou:1991cr, BieriChenYau2011}.

Additionally, our reconstruction procedure could help extend the modified Teukolsky formalism
\cite{Li:2022pcy,Cano:2023tmv} to charged, rotating black holes with
higher-derivative corrections. This program builds on renewed interest
in gravitoelectromagnetic perturbations in higher-derivative theories
and nonlinear electrodynamics
\cite{Chaverra:2016ttw,Nomura:2021efi,Boyce:2025fpr,Ji:2026qeo,
Stashko:2026kja}, together with studies of corrected charged black-hole
backgrounds \cite{Ma:2024ulp,Ajith:2026eat,Cano:2019ycn,Verbin:2020fzk}. Such studies could allow future work that examines if isospectrality is preserved in modified Einstein--Maxwell theories. 

More broadly, this opens up new directions for studying connections
between the classical scattering matrix of rotating, charged black
holes in asymptotically de~Sitter space and their radiative geometry.
The reconstructed fields could also be used to investigate how bulk
dynamics are imprinted on the future black-hole and cosmological
horizons, with potential applications to de~Sitter holography
\cite{Anninos:2009yc, Anninos:2010gh}.

\textit{Acknowledgments}---We thank Hector Silva for interesting discussions. C.~W., A.~L., and Y.~C.'s research is supported by the Simons Foundation (Award No. 568762),
the Brinson Foundation, and the National Science Foundation (via Grants No. PHY-2011961 and No. PHY-2011968). A.L. acknowledges support from the Fannie and John Hertz Foundation in the form of a Hertz Fellowship. D.~L. and N.~Y. acknowledge support from the Simons Foundation (via Award No. 896696), the Simons Foundation International (via Grant No. SFI-MPS-BH-00012593-01), and the NSF (via Grants No.~PHY-2512423). P.~W. acknowledges funding from the Deutsche Forschungsgemeinschaft (DFG) project number: 386119226.

\bibliographystyle{apsrev4-1}
\bibliography{reference}

@article{carter1968new,
  title={A new family of Einstein spaces},
  author={Carter, Brandon},
  journal={Physics Letters A},
  volume={26},
  number={9},
  pages={399--400},
  year={1968},
  publisher={Elsevier}
}

@article{Weller:2026jqu,
    author = "Weller, Colin and Li, Dongjun and Wagle, Pratik and Laeuger, Andrew and Chen, Yanbei and Yunes, Nicol{\'a}s",
    title = "{Chiral symmetry and black hole isospectrality}",
    eprint = "2608.21532",
    archivePrefix = "arXiv",
    primaryClass = "gr-qc",
    month = "8",
    year = "2026"
}

@article{Suzuki:1998vy,
    author = "Suzuki, Hisao and Takasugi, Eiichi and Umetsu, Hiroshi",
    title = "{Perturbations of Kerr-de Sitter black hole and Heun's equations}",
    eprint = "gr-qc/9805064",
    archivePrefix = "arXiv",
    reportNumber = "EPHOU-98-005, OU-HET-296",
    doi = "10.1143/PTP.100.491",
    journal = "Prog. Theor. Phys.",
    volume = "100",
    pages = "491--505",
    year = "1998"
}

@misc{SupplementalMaterial,
  title        = {Supplemental Material for ``Parity Isospectrality of the Kerr--Newman de--Sitter Black Hole''},
}

@article{Bah:2026aia,
    author = "Bah, Ibrahima and Berti, Emanuele and De Luca, Valerio and Ganchev, Bogdan and Pere{\~n}iguez, David",
    title = "{Gravitational electric-magnetic duality at the light ring and quasinormal mode isospectrality in effective field theories}",
    eprint = "2605.03018",
    archivePrefix = "arXiv",
    primaryClass = "gr-qc",
    month = "5",
    year = "2026"
}

@article{Pani:2013ija,
    author = "Pani, Paolo and Berti, Emanuele and Gualtieri, Leonardo",
    title = "{Gravitoelectromagnetic Perturbations of Kerr-Newman Black Holes: Stability and Isospectrality in the Slow-Rotation Limit}",
    eprint = "1304.1160",
    archivePrefix = "arXiv",
    primaryClass = "gr-qc",
    doi = "10.1103/PhysRevLett.110.241103",
    journal = "Phys. Rev. Lett.",
    volume = "110",
    number = "24",
    pages = "241103",
    year = "2013"
}

@article{held1974formalism,
  title={A formalism for the investigation of algebraically special metrics. I},
  author={Held, Alan},
  journal={Communications in Mathematical Physics},
  volume={37},
  number={4},
  pages={311--326},
  year={1974},
  publisher={Springer}
}

@article{Berens:2025kkm,
    author = "Berens, Roman and Gravely, Trevor and Lupsasca, Alexandru",
    title = "{Gravitational waves on Kerr black holes: II. Metric reconstruction with cosmological constant}",
    eprint = "2510.07712",
    archivePrefix = "arXiv",
    primaryClass = "gr-qc",
    doi = "10.1088/1361-6382/ae4282",
    journal = "Class. Quant. Grav.",
    volume = "43",
    number = "4",
    pages = "045007",
    year = "2026"
}

@article{Chambers:1994ap,
    author = "Chambers, Chris M. and Moss, Ian G.",
    title = "{Stability of the Cauchy horizon in Kerr-de Sitter space-times}",
    eprint = "gr-qc/9404015",
    archivePrefix = "arXiv",
    reportNumber = "NCL-93-TP-10",
    doi = "10.1088/0264-9381/11/4/019",
    journal = "Class. Quant. Grav.",
    volume = "11",
    pages = "1035--1054",
    year = "1994"
}

@article{Dias:2013sdc,
    author = "Dias, {\'O}scar J. C. and Santos, Jorge E.",
    title = "{Boundary Conditions for Kerr-AdS Perturbations}",
    eprint = "1302.1580",
    archivePrefix = "arXiv",
    primaryClass = "hep-th",
    doi = "10.1007/JHEP10(2013)156",
    journal = "JHEP",
    volume = "10",
    pages = "156",
    year = "2013"
}

@article{Chandrasekhar:1979iz,
    author = "Chandrasekhar, Subrahmanyan",
    editor = "Wali, K. C.",
    title = "{ON THE EQUATIONS GOVERNING THE PERTURBATIONS OF THE REISSNER-NORDSTROM BLACK HOLE}",
    doi = "10.1098/rspa.1979.0028",
    journal = "Proc. Roy. Soc. Lond. A",
    volume = "365",
    pages = "453--465",
    year = "1979"
}

@article{Dias:2015wqa,
    author = "Dias, Oscar J. C. and Godazgar, Mahdi and Santos, Jorge E.",
    title = "{Linear Mode Stability of the Kerr-Newman Black Hole and Its Quasinormal Modes}",
    eprint = "1501.04625",
    archivePrefix = "arXiv",
    primaryClass = "gr-qc",
    doi = "10.1103/PhysRevLett.114.151101",
    journal = "Phys. Rev. Lett.",
    volume = "114",
    number = "15",
    pages = "151101",
    year = "2015"
}

@article{Price:2006ke,
    author = "Price, Larry R. and Shankar, Karthik and Whiting, Bernard F.",
    title = "{On the existence of radiation gauges in Petrov type II spacetimes}",
    eprint = "gr-qc/0611070",
    archivePrefix = "arXiv",
    doi = "10.1088/0264-9381/24/9/014",
    journal = "Class. Quant. Grav.",
    volume = "24",
    pages = "2367--2388",
    year = "2007"
}

@article{Li:2023ulk,
    author = "Li, Dongjun and Hussain, Asad and Wagle, Pratik and Chen, Yanbei and Yunes, Nicol\'as and Zimmerman, Aaron",
    title = "{Isospectrality breaking in the Teukolsky formalism}",
    eprint = "2310.06033",
    archivePrefix = "arXiv",
    primaryClass = "gr-qc",
    reportNumber = "UTWI-14-2023",
    doi = "10.1103/PhysRevD.109.104026",
    journal = "Phys. Rev. D",
    volume = "109",
    number = "10",
    pages = "104026",
    year = "2024"
}

@article{Wagle:2023fwl,
    author = "Wagle, Pratik and Li, Dongjun and Chen, Yanbei and Yunes, Nicolas",
    title = "{Perturbations of spinning black holes in dynamical Chern-Simons gravity: Slow rotation equations}",
    eprint = "2311.07706",
    archivePrefix = "arXiv",
    primaryClass = "gr-qc",
    doi = "10.1103/PhysRevD.109.104029",
    journal = "Phys. Rev. D",
    volume = "109",
    number = "10",
    pages = "104029",
    year = "2024"
}

@article{Mei:2026oem,
    author = "Mei, Jianwei",
    title = "{Separating the linearized Einstein equations in Kerr}",
    eprint = "2607.16581",
    archivePrefix = "arXiv",
    primaryClass = "gr-qc",
    month = "7",
    year = "2026"
}

@article{Mei:2023pho,
    author = "Mei, Jianwei",
    title = "{Separating the linearized Einstein equations in the background of a Kerr black hole}",
    eprint = "2311.18409",
    archivePrefix = "arXiv",
    primaryClass = "gr-qc",
    doi = "10.1007/s11433-026-2975-2",
    journal = "Sci. China Phys. Mech. Astron.",
    volume = "69",
    number = "7",
    pages = "270415",
    year = "2026"
}

@article{de1987gravitational,
  title={Gravitational and electromagnetic perturbations of the Reissner-Nordstrom solution},
  author={de Castillo, GF Torres},
  journal={Classical and Quantum Gravity},
  volume={4},
  number={5},
  pages={1133--1142},
  year={1987}
}

@article{torres1987complex,
  title={Complex potentials for coupled gravitational and electromagnetic perturbations},
  author={Torres del Castillo, GF},
  journal={Classical and Quantum Gravity},
  volume={4},
  number={4},
  pages={1011--1017},
  year={1987}
}

@article{dudley1979covariant,
  title={Covariant perturbed wave equations in arbitrary type-D backgrounds},
  author={Dudley, Alan L and Finley III, JD},
  journal={Journal of Mathematical Physics},
  volume={20},
  number={2},
  pages={311--328},
  year={1979},
  publisher={American Institute of Physics}
}

@article{Saha:2025nsg,
    author = "Saha, Sagnik and Silva, Hector O.",
    title = "{Quasinormal modes of Kerr-Newman black holes: Revisiting the Dudley-Finley approximation}",
    eprint = "2510.05354",
    archivePrefix = "arXiv",
    primaryClass = "gr-qc",
    doi = "10.1103/2wc1-yntl",
    journal = "Phys. Rev. D",
    volume = "113",
    number = "6",
    pages = "064009",
    year = "2026"
}

@article{Dudley:1977zz,
    author = "Dudley, Alan L. and Finley, J. D.",
    title = "{Separation of Wave Equations for Perturbations of General Type-D Space-Times}",
    doi = "10.1103/PhysRevLett.38.1505",
    journal = "Phys. Rev. Lett.",
    volume = "38",
    pages = "1505--1508",
    year = "1977"
}

@article{Casals:2005kr,
    author = "Casals, Marc and Ottewill, Adrian C.",
    title = "{Canonical quantization of the electromagnetic field on the Kerr background}",
    eprint = "gr-qc/0501005",
    archivePrefix = "arXiv",
    doi = "10.1103/PhysRevD.71.124016",
    journal = "Phys. Rev. D",
    volume = "71",
    pages = "124016",
    year = "2005"
}

@article{Miguel:2023rzp,
    author = "Miguel, Filipe S.",
    title = "{EFT corrections to scalar and vector quasinormal modes of rapidly rotating black holes}",
    eprint = "2308.03832",
    archivePrefix = "arXiv",
    primaryClass = "gr-qc",
    doi = "10.1103/PhysRevD.109.104016",
    journal = "Phys. Rev. D",
    volume = "109",
    number = "10",
    pages = "104016",
    year = "2024"
}

@article{Loutrel:2020wbw,
    author = "Loutrel, Nicholas and Ripley, Justin L. and Giorgi, Elena and Pretorius, Frans",
    title = "{Second Order Perturbations of Kerr Black Holes: Reconstruction of the Metric}",
    eprint = "2008.11770",
    archivePrefix = "arXiv",
    primaryClass = "gr-qc",
    doi = "10.1103/PhysRevD.103.104017",
    journal = "Phys. Rev. D",
    volume = "103",
    number = "10",
    pages = "104017",
    year = "2021"
}

@article{Wald:1978vm,
    author = "Wald, Robert M.",
    title = "{Construction of Solutions of Gravitational, Electromagnetic, Or Other Perturbation Equations from Solutions of Decoupled Equations}",
    doi = "10.1103/PhysRevLett.41.203",
    journal = "Phys. Rev. Lett.",
    volume = "41",
    pages = "203--206",
    year = "1978"
}

@article{Chrzanowski:1976jy,
    author = "Chrzanowski, P. L.",
    title = "{Applications of Metric Perturbations of a Rotating Black Hole: Distortion of the Event Horizon}",
    doi = "10.1103/PhysRevD.13.806",
    journal = "Phys. Rev. D",
    volume = "13",
    pages = "806--818",
    year = "1976"
}

@article{Li:2022pcy,
    author = "Li, Dongjun and Wagle, Pratik and Chen, Yanbei and Yunes, Nicol\'as",
    title = "{Perturbations of Spinning Black Holes beyond General Relativity: Modified Teukolsky Equation}",
    eprint = "2206.10652",
    archivePrefix = "arXiv",
    primaryClass = "gr-qc",
    doi = "10.1103/PhysRevX.13.021029",
    journal = "Phys. Rev. X",
    volume = "13",
    number = "2",
    pages = "021029",
    year = "2023"
}

@article{Cohen_Kegeles_1975, 
    title={Space-time perturbations}, 
    volume={54}, 
    ISSN={0375-9601}, 
    DOI={10.1016/0375-9601(75)90583-6}, 
    number={1}, 
    journal={Physics Letters A}, 
    author={Cohen, J. M. and Kegeles, L. S.}, 
    year={1975}, month={Aug}, 
    pages={5-7}
}

@article{Nichols:2012jn,
    author = "Nichols, David A. and Zimmerman, Aaron and Chen, Yanbei and Lovelace, Geoffrey and Matthews, Keith D. and Owen, Robert and Zhang, Fan and Thorne, Kip S.",
    title = "{Visualizing Spacetime Curvature via Frame-Drag Vortexes and Tidal Tendexes III. Quasinormal Pulsations of Schwarzschild and Kerr Black Holes}",
    eprint = "1208.3038",
    archivePrefix = "arXiv",
    primaryClass = "gr-qc",
    doi = "10.1103/PhysRevD.86.104028",
    journal = "Phys. Rev. D",
    volume = "86",
    pages = "104028",
    year = "2012"
}

@article{Petrov:2000bs,
    author = "Petrov, A. Z.",
    title = "{The Classification of spaces defining gravitational fields}",
    doi = "10.1023/A:1001910908054",
    journal = "Gen. Rel. Grav.",
    volume = "32",
    pages = "1661--1663",
    year = "2000"
}

@article{Mark:2014aja,
    author = "Mark, Zachary and Yang, Huan and Zimmerman, Aaron and Chen, Yanbei",
    title = "{Quasinormal modes of weakly charged Kerr-Newman spacetimes}",
    eprint = "1409.5800",
    archivePrefix = "arXiv",
    primaryClass = "gr-qc",
    doi = "10.1103/PhysRevD.91.044025",
    journal = "Phys. Rev. D",
    volume = "91",
    number = "4",
    pages = "044025",
    year = "2015"
}

@article{Kegeles:1979an,
    author = "Kegeles, L. S. and Cohen, J. M.",
    title = "{CONSTRUCTIVE PROCEDURE FOR PERTURBATIONS OF SPACE-TIMES}",
    doi = "10.1103/PhysRevD.19.1641",
    journal = "Phys. Rev. D",
    volume = "19",
    pages = "1641--1664",
    year = "1979"
}

@article{Whiting:2005hr,
    author = "Whiting, B. F. and Price, L. R.",
    title = "{Metric reconstruction from Weyl scalars}",
    doi = "10.1088/0264-9381/22/15/003",
    journal = "Class. Quant. Grav.",
    volume = "22",
    pages = "S589--S604",
    year = "2005"
}

@article{Newman:1961qr,
    author = "Newman, Ezra and Penrose, Roger",
    title = "{An Approach to gravitational radiation by a method of spin coefficients}",
    doi = "10.1063/1.1724257",
    journal = "J. Math. Phys.",
    volume = "3",
    pages = "566--578",
    year = "1962"
}

@article{Teukolsky:1974yv,
    author = "Teukolsky, S. A. and Press, W. H.",
    title = "{Perturbations of a rotating black hole. III - Interaction of the hole with gravitational and electromagnetic radiation}",
    doi = "10.1086/153180",
    journal = "Astrophys. J.",
    volume = "193",
    pages = "443--461",
    year = "1974"
}

@article{Aksteiner:2022bwr,
    author = "Aksteiner, Steffen and Araneda, Bernardo",
    title = {{K{\"a}hler Geometry of Black Holes and Gravitational Instantons}},
    eprint = "2207.10039",
    archivePrefix = "arXiv",
    primaryClass = "gr-qc",
    doi = "10.1103/PhysRevLett.130.161502",
    journal = "Phys. Rev. Lett.",
    volume = "130",
    number = "16",
    pages = "161502",
    year = "2023"
}

@article{Newman:1965my,
    author = "Newman, E T. and Couch, R. and Chinnapared, K. and Exton, A. and Prakash, A. and Torrence, R.",
    title = "{Metric of a Rotating, Charged Mass}",
    doi = "10.1063/1.1704351",
    journal = "J. Math. Phys.",
    volume = "6",
    pages = "918--919",
    year = "1965"
}

@article{Regge:1957td,
    author = "Regge, Tullio and Wheeler, John A.",
    title = "{Stability of a Schwarzschild singularity}",
    doi = "10.1103/PhysRev.108.1063",
    journal = "Phys. Rev.",
    volume = "108",
    pages = "1063--1069",
    year = "1957"
}

@article{Chrzanowski:1975wv,
    author = "Chrzanowski, P. L.",
    title = "{Vector Potential and Metric Perturbations of a Rotating Black Hole}",
    doi = "10.1103/PhysRevD.11.2042",
    journal = "Phys. Rev. D",
    volume = "11",
    pages = "2042--2062",
    year = "1975"
}

@article{Chandrasekhar:1975zza,
    author = "Chandrasekhar, S. and Detweiler, Steven L.",
    title = "{The quasi-normal modes of the Schwarzschild black hole}",
    doi = "10.1098/rspa.1975.0112",
    journal = "Proc. Roy. Soc. Lond. A",
    volume = "344",
    pages = "441--452",
    year = "1975"
}

@article{Anderson:1991kx,
    author = "Anderson, A. and Price, R. H.",
    title = "{Intertwining of the equations of black hole perturbations}",
    doi = "10.1103/PhysRevD.43.3147",
    journal = "Phys. Rev. D",
    volume = "43",
    pages = "3147--3154",
    year = "1991"
}

@article{Blazquez-Salcedo:2022eik,
    author = "Bl{\'a}zquez-Salcedo, Jose Luis and Khoo, Fech Scen",
    title = "{Quasinormal modes of slowly rotating Kerr-Newman black holes using the double series method}",
    eprint = "2212.00054",
    archivePrefix = "arXiv",
    primaryClass = "gr-qc",
    doi = "10.1103/PhysRevD.107.084031",
    journal = "Phys. Rev. D",
    volume = "107",
    number = "8",
    pages = "084031",
    year = "2023"
}

@article{Teukolsky:1973ha,
    author = "Teukolsky, Saul A.",
    title = "{Perturbations of a rotating black hole. 1. Fundamental equations for gravitational electromagnetic and neutrino field perturbations}",
    doi = "10.1086/152444",
    journal = "Astrophys. J.",
    volume = "185",
    pages = "635--647",
    year = "1973"
}

@article{Teukolsky:1972my,
    author = "Teukolsky, S. A.",
    title = "{Rotating black holes - separable wave equations for gravitational and electromagnetic perturbations}",
    reportNumber = "OAP-291",
    doi = "10.1103/PhysRevLett.29.1114",
    journal = "Phys. Rev. Lett.",
    volume = "29",
    pages = "1114--1118",
    year = "1972"
}

@article{Cano:2023tmv,
    author = "Cano, Pablo A. and Fransen, Kwinten and Hertog, Thomas and Maenaut, Simon",
    title = "{Universal Teukolsky equations and black hole perturbations in higher-derivative gravity}",
    eprint = "2304.02663",
    archivePrefix = "arXiv",
    primaryClass = "gr-qc",
    doi = "10.1103/PhysRevD.108.024040",
    journal = "Phys. Rev. D",
    volume = "108",
    number = "2",
    pages = "024040",
    year = "2023"
}

@article{Moncrief:1974am,
    author = "Moncrief, V.",
    title = "{Gravitational perturbations of spherically symmetric systems. I. The exterior problem.}",
    doi = "10.1016/0003-4916(74)90173-0",
    journal = "Annals Phys.",
    volume = "88",
    pages = "323--342",
    year = "1974"
}

@article{Li:2026rkf,
    author = "Li, Dongjun and Yunes, Nicol{\'a}s",
    title = "{Metric Reconstruction for Generic Black-Hole Perturbations}",
    eprint = "2605.11080",
    archivePrefix = "arXiv",
    primaryClass = "gr-qc",
    month = "5",
    year = "2026"
}

@article{Mellor:1989ac,
    author = "Mellor, Felicity and Moss, Ian",
    title = "{Stability of Black Holes in De Sitter Space}",
    reportNumber = "NCL-89-TP10",
    doi = "10.1103/PhysRevD.41.403",
    journal = "Phys. Rev. D",
    volume = "41",
    pages = "403",
    year = "1990"
}

@article{Zerilli:1970wzz,
    author = "Zerilli, F. J.",
    title = "{Gravitational field of a particle falling in a schwarzschild geometry analyzed in tensor harmonics}",
    doi = "10.1103/PhysRevD.2.2141",
    journal = "Phys. Rev. D",
    volume = "2",
    pages = "2141--2160",
    year = "1970"
}

@article{Boyeneni:2025tsx,
    author = "Boyeneni, Siddharth and Wu, Jiaxi and Most, Elias R.",
    title = "{Unveiling the Electrodynamic Nature of Spacetime Collisions}",
    eprint = "2504.15978",
    archivePrefix = "arXiv",
    primaryClass = "gr-qc",
    doi = "10.1103/995s-wxl7",
    journal = "Phys. Rev. Lett.",
    volume = "135",
    number = "10",
    pages = "101401",
    year = "2025"
}

@article{Zhang:2012jj,
    author = "Zhang, Fan and Zimmerman, Aaron and Nichols, David A. and Chen, Yanbei and Lovelace, Geoffrey and Matthews, Keith D. and Owen, Robert and Thorne, Kip S.",
    title = "{Visualizing Spacetime Curvature via Frame-Drag Vortexes and Tidal Tendexes II. Stationary Black Holes}",
    eprint = "1208.3034",
    archivePrefix = "arXiv",
    primaryClass = "gr-qc",
    doi = "10.1103/PhysRevD.86.084049",
    journal = "Phys. Rev. D",
    volume = "86",
    pages = "084049",
    year = "2012"
}

@article{Nichols:2011pu,
    author = "Nichols, David A. and others",
    title = "{Visualizing Spacetime Curvature via Frame-Drag Vortexes and Tidal Tendexes I. General Theory and Weak-Gravity Applications}",
    eprint = "1108.5486",
    archivePrefix = "arXiv",
    primaryClass = "gr-qc",
    doi = "10.1103/PhysRevD.84.124014",
    journal = "Phys. Rev. D",
    volume = "84",
    pages = "124014",
    year = "2011"
}

@article{GuvendeSitter,
  title = {Schwarzschild-de Sitter space and its perturbations},
  author = {Guven, Jemal and N\'u\~nez, Dar\'{\i}o},
  journal = {Phys. Rev. D},
  volume = {42},
  issue = {8},
  pages = {2577--2584},
  numpages = {0},
  year = {1990},
  month = {Oct},
  publisher = {American Physical Society},
  doi = {10.1103/PhysRevD.42.2577},
  url = {https://link.aps.org/doi/10.1103/PhysRevD.42.2577}
}

@article{bunting1983proof,
  title={Proof of the uniqueness conjecture for black holes},
  author={Bunting, GL},
  journal={PhD Thesis, University of New England},
  year={1983}
}

@article{Mazur:1982db,
    author = "Mazur, P. O.",
    title = "{PROOF OF UNIQUENESS OF THE KERR-NEWMAN BLACK HOLE SOLUTION}",
    doi = "10.1088/0305-4470/15/10/021",
    journal = "J. Phys. A",
    volume = "15",
    pages = "3173--3180",
    year = "1982"
}

@book{Chandrasekhar_1983, 
	place={Oxford [Oxfordshire]: New York}, 
 	series={The International series of monographs on physics}, 
 	title={The mathematical theory of black holes}, 
 	ISBN={978-0-19-851291-2}, 
 	publisher={Clarendon Press}, 
 	author={Chandrasekhar, S.}, 
 	year={1983}, 
 	collection={The International series of monographs on physics}
}

@article{Khanal:1983kk,
  title = {Rotating black hole in asymptotic de Sitter space: Perturbation of the space-time with spin fields},
  author = {Khanal, U.},
  journal = {Phys. Rev. D},
  volume = {28},
  issue = {6},
  pages = {1291--1297},
  numpages = {0},
  year = {1983},
  month = {Sep},
  publisher = {American Physical Society},
  doi = {10.1103/PhysRevD.28.1291},
  url = {https://link.aps.org/doi/10.1103/PhysRevD.28.1291}
}

@article{Zerilli:1974ai,
    author = "Zerilli, F. J.",
    title = "{Perturbation analysis for gravitational and electromagnetic radiation in a reissner-nordstroem geometry}",
    doi = "10.1103/PhysRevD.9.860",
    journal = "Phys. Rev. D",
    volume = "9",
    pages = "860--868",
    year = "1974"
}

@article{Campanelli:1998jv,
    author = "Campanelli, Manuela and Lousto, Carlos O.",
    title = "{Second order gauge invariant gravitational perturbations of a Kerr black hole}",
    eprint = "gr-qc/9811019",
    archivePrefix = "arXiv",
    reportNumber = "AEI-096",
    doi = "10.1103/PhysRevD.59.124022",
    journal = "Phys. Rev. D",
    volume = "59",
    pages = "124022",
    year = "1999"
}

@article{Green:2019nam,
    author = "Green, Stephen R. and Hollands, Stefan and Zimmerman, Peter",
    title = "{Teukolsky formalism for nonlinear Kerr perturbations}",
    eprint = "1908.09095",
    archivePrefix = "arXiv",
    primaryClass = "gr-qc",
    doi = "10.1088/1361-6382/ab7075",
    journal = "Class. Quant. Grav.",
    volume = "37",
    number = "7",
    pages = "075001",
    year = "2020"
}

@article{Aksteiner:2016pjt,
    author = {Aksteiner, Steffen and Andersson, Lars and B{\"a}ckdahl, Thomas},
    title = "{New identities for linearized gravity on the Kerr spacetime}",
    eprint = "1601.06084",
    archivePrefix = "arXiv",
    primaryClass = "gr-qc",
    doi = "10.1103/PhysRevD.99.044043",
    journal = "Phys. Rev. D",
    volume = "99",
    number = "4",
    pages = "044043",
    year = "2019"
}

@article{Ma:2024qcv,
    author = "Ma, Sizheng and Yang, Huan",
    title = "{Excitation of quadratic quasinormal modes for Kerr black holes}",
    eprint = "2401.15516",
    archivePrefix = "arXiv",
    primaryClass = "gr-qc",
    doi = "10.1103/PhysRevD.109.104070",
    journal = "Phys. Rev. D",
    volume = "109",
    number = "10",
    pages = "104070",
    year = "2024"
}

@article{Pani:2013wsa,
    author = "Pani, Paolo and Berti, Emanuele and Gualtieri, Leonardo",
    title = "{Scalar, Electromagnetic and Gravitational Perturbations of Kerr-Newman Black Holes in the Slow-Rotation Limit}",
    eprint = "1307.7315",
    archivePrefix = "arXiv",
    primaryClass = "gr-qc",
    doi = "10.1103/PhysRevD.88.064048",
    journal = "Phys. Rev. D",
    volume = "88",
    pages = "064048",
    year = "2013"
}

@article{Press:1973zz,
    author = "Press, William H. and Teukolsky, Saul A.",
    title = "{Perturbations of a Rotating Black Hole. II. Dynamical Stability of the Kerr Metric}",
    doi = "10.1086/152445",
    journal = "Astrophys. J.",
    volume = "185",
    pages = "649--674",
    year = "1973"
}

@book{Wheeler:1962kl,
  author    = {Wheeler, John Archibald},
  title     = {Geometrodynamics},
  publisher = {Academic Press},
  address   = {New York},
  year      = {1962}
}

@article{Glampedakis:2017rar,
    author = "Glampedakis, Kostas and Johnson, Aaron D. and Kennefick, Daniel",
    title = "{Darboux transformation in black hole perturbation theory}",
    eprint = "1702.06459",
    archivePrefix = "arXiv",
    primaryClass = "gr-qc",
    doi = "10.1103/PhysRevD.96.024036",
    journal = "Phys. Rev. D",
    volume = "96",
    number = "2",
    pages = "024036",
    year = "2017"
}

@article{Pereniguez:2026avs,
    author = "Pere{\~n}iguez, David",
    title = "{Unifying the Regge-Wheeler-Zerilli and Bardeen-Press-Teukolsky formalisms on spherical backgrounds}",
    eprint = "2605.04147",
    archivePrefix = "arXiv",
    primaryClass = "gr-qc",
    month = "5",
    year = "2026"
}

@article{Bourg:2024vre,
    author = "Bourg, Patrick and Leather, Benjamin and Casals, Marc and Pound, Adam and Wardell, Barry",
    title = "{Implementation of a Green-Hollands-Zimmerman-Teukolsky puncture scheme for gravitational self-force calculations}",
    eprint = "2403.12634",
    archivePrefix = "arXiv",
    primaryClass = "gr-qc",
    doi = "10.1103/PhysRevD.110.044007",
    journal = "Phys. Rev. D",
    volume = "110",
    number = "4",
    pages = "044007",
    year = "2024"
}

@article{vandeMeent:2017bcc,
    author = "van de Meent, Maarten",
    title = "{Gravitational self-force on generic bound geodesics in Kerr spacetime}",
    eprint = "1711.09607",
    archivePrefix = "arXiv",
    primaryClass = "gr-qc",
    doi = "10.1103/PhysRevD.97.104033",
    journal = "Phys. Rev. D",
    volume = "97",
    number = "10",
    pages = "104033",
    year = "2018"
}

@article{Wardell:2020naz,
    author = "Wardell, Barry and Kavanagh, Chris",
    title = "{Separable electromagnetic perturbations of rotating black holes}",
    eprint = "2011.03548",
    archivePrefix = "arXiv",
    primaryClass = "gr-qc",
    doi = "10.1103/PhysRevD.103.104049",
    journal = "Phys. Rev. D",
    volume = "103",
    number = "10",
    pages = "104049",
    year = "2021"
}

@article{Boyer:1966qh,
    author = "Boyer, Robert H. and Lindquist, Richard W.",
    title = "{Maximal analytic extension of the Kerr metric}",
    doi = "10.1063/1.1705193",
    journal = "J. Math. Phys.",
    volume = "8",
    pages = "265",
    year = "1967"
}

@article{Geroch:1973am,
    author = "Geroch, Robert P. and Held, A. and Penrose, R.",
    title = "{A space-time calculus based on pairs of null directions}",
    doi = "10.1063/1.1666410",
    journal = "J. Math. Phys.",
    volume = "14",
    pages = "874--881",
    year = "1973"
}

@article{Zerilli:1970se,
    author = "Zerilli, Frank J.",
    title = "{Effective potential for even parity Regge-Wheeler gravitational perturbation equations}",
    doi = "10.1103/PhysRevLett.24.737",
    journal = "Phys. Rev. Lett.",
    volume = "24",
    pages = "737--738",
    year = "1970"
}

@article{Kraniotis:2018zmh,
    author = "Kraniotis, G. V.",
    title = "{The massive Dirac equation in the Kerr-Newman-de Sitter and Kerr-Newman black hole spacetimes}",
    eprint = "1801.03157",
    archivePrefix = "arXiv",
    primaryClass = "gr-qc",
    doi = "10.1088/2399-6528/ab1046",
    journal = "J. Phys. Comm.",
    volume = "3",
    pages = "035026",
    year = "2019"
}

@article{Toomani:2021jlo,
    author = "Toomani, Vahid and Zimmerman, Peter and Spiers, Andrew and Hollands, Stefan and Pound, Adam and Green, Stephen R.",
    title = "{New metric reconstruction scheme for gravitational self-force calculations}",
    eprint = "2108.04273",
    archivePrefix = "arXiv",
    primaryClass = "gr-qc",
    doi = "10.1088/1361-6382/ac37a5",
    journal = "Class. Quant. Grav.",
    volume = "39",
    number = "1",
    pages = "015019",
    year = "2022"
}

@article{lee1976coupled,
  title={Coupled gravitational and electromagnetic perturbations around a charged black hole},
  author={Lee, Chul Hoon},
  journal={Journal of Mathematical Physics},
  volume={17},
  number={7},
  pages={1226--1235},
  year={1976},
  publisher={American Institute of Physics}
}

@article{Boyce:2025fpr,
    author = "Boyce, William L. and Santos, Jorge E.",
    title = "{EFT corrections to charged black hole quasinormal modes}",
    eprint = "2506.10074",
    archivePrefix = "arXiv",
    primaryClass = "hep-th",
    doi = "10.1007/JHEP12(2025)017",
    journal = "JHEP",
    volume = "12",
    pages = "017",
    year = "2025"
}

@article{Kokkotas:1993ef,
    author = "Kokkotas, K. D.",
    title = "{Quasinormal modes of the Kerr-Newman black hole}",
    doi = "10.1007/BF02822861",
    journal = "Nuovo Cim. B",
    volume = "108",
    pages = "991--998",
    year = "1993"
}

@article{Berti:2005eb,
    author = "Berti, Emanuele and Kokkotas, Kostas D.",
    title = "{Quasinormal modes of Kerr-Newman black holes: Coupling of electromagnetic and gravitational perturbations}",
    eprint = "gr-qc/0502065",
    archivePrefix = "arXiv",
    doi = "10.1103/PhysRevD.71.124008",
    journal = "Phys. Rev. D",
    volume = "71",
    pages = "124008",
    year = "2005"
}

@article{Verbin:2020fzk,
    author = "Verbin, Y.",
    title = "{Magnetic and electric black holes in the vector-tensor Horndeski theory}",
    eprint = "2011.02515",
    archivePrefix = "arXiv",
    primaryClass = "gr-qc",
    doi = "10.1103/PhysRevD.106.024057",
    journal = "Phys. Rev. D",
    volume = "106",
    number = "2",
    pages = "024057",
    year = "2022"
}

@article{Cano:2019ycn,
    author = "Cano, Pablo A. and Chimento, Samuele and Linares, Rom{\'a}n and Ort{\'\i}n, Tom{\'a}s and Ram{\'\i}rez, Pedro F.",
    title = {{$\alpha'$ corrections of Reissner-Nordstr{\"o}m black holes}},
    eprint = "1910.14324",
    archivePrefix = "arXiv",
    primaryClass = "hep-th",
    reportNumber = "IFT-UAM/CSIC-18-114",
    doi = "10.1007/JHEP02(2020)031",
    journal = "JHEP",
    volume = "02",
    pages = "031",
    year = "2020"
}

@article{Ajith:2026eat,
    author = "Ajith, Siddarth and Rajagopal, Ravisankar and Rifat, Nur and Vaman, Diana and Yagi, Kent",
    title = {{Higher-Derivative Corrections to Reissner--Nordstr{\"o}m Black Holes from Worldline QFT}},
    eprint = "2605.31331",
    archivePrefix = "arXiv",
    primaryClass = "hep-th",
    month = "5",
    year = "2026"
}

@article{Ma:2024ulp,
    author = "Ma, Liang and Pang, Yi and Lu, H.",
    title = "{Leading higher derivative corrections to multipole moments of Kerr-Newman black hole}",
    eprint = "2411.13639",
    archivePrefix = "arXiv",
    primaryClass = "hep-th",
    reportNumber = "USTC-ICTS/PCFT-24-61",
    doi = "10.1007/JHEP02(2025)079",
    journal = "JHEP",
    volume = "02",
    pages = "079",
    year = "2025"
}

@article{Chaverra:2016ttw,
    author = "Chaverra, Eliana and Degollado, Juan Carlos and Moreno, Claudia and Sarbach, Olivier",
    title = "{Black holes in nonlinear electrodynamics: Quasinormal spectra and parity splitting}",
    eprint = "1605.04003",
    archivePrefix = "arXiv",
    primaryClass = "gr-qc",
    doi = "10.1103/PhysRevD.93.123013",
    journal = "Phys. Rev. D",
    volume = "93",
    number = "12",
    pages = "123013",
    year = "2016"
}

@article{Stashko:2026kja,
    author = "Stashko, Oleksandr S. and Konoplya, Roman A. and Hertzberg, Mark P.",
    title = "{Perturbations of Charged Black Holes with Higher-Order Interactions}",
    eprint = "2609.09419",
    archivePrefix = "arXiv",
    primaryClass = "gr-qc",
    month = "9",
    year = "2026"
}

@article{Ji:2026qeo,
    author = "Ji, Yuanyuan and Long, Zheng-Wen",
    title = "{Gravito-Electromagnetic perturbation of higher derivative corrected Kerr-Newman black hole}",
    eprint = "2607.08333",
    archivePrefix = "arXiv",
    primaryClass = "gr-qc",
    month = "7",
    year = "2026"
}

@article{Nomura:2021efi,
    author = "Nomura, Kimihiro and Yoshida, Daisuke",
    title = "{Quasinormal modes of charged black holes with corrections from nonlinear electrodynamics}",
    eprint = "2111.06273",
    archivePrefix = "arXiv",
    primaryClass = "gr-qc",
    reportNumber = "KOBE-COSMO-21-16",
    doi = "10.1103/PhysRevD.105.044006",
    journal = "Phys. Rev. D",
    volume = "105",
    number = "4",
    pages = "044006",
    year = "2022"
}

@article{ZimmermanPoisson2014,
  author        = {Zimmerman, Peter and Poisson, Eric},
  title         = {Gravitational self-force in nonvacuum spacetimes},
  journal       = {Phys. Rev. D},
  volume        = {90},
  pages         = {084030},
  year          = {2014},
  doi           = {10.1103/PhysRevD.90.084030},
  eprint        = {1406.5111},
  archivePrefix = {arXiv},
  primaryClass  = {gr-qc}
}

@article{BieriChenYau2011,
  author        = {Bieri, Lydia and Chen, PoNing and Yau, Shing-Tung},
  title         = {Null Asymptotics of Solutions of the
                   {Einstein--Maxwell} Equations in General Relativity
                   and Gravitational Radiation},
  journal       = {Adv. Theor. Math. Phys.},
  volume        = {15},
  number        = {4},
  pages         = {1085--1113},
  year          = {2011},
  doi           = {10.4310/ATMP.2011.v15.n4.a5},
  eprint        = {1011.2267},
  archivePrefix = {arXiv},
  primaryClass  = {math.DG}
}

@article{Anninos:2010gh,
    author = "Anninos, Dionysios and Anous, Tarek",
    title = "{A de Sitter Hoedown}",
    eprint = "1002.1717",
    archivePrefix = "arXiv",
    primaryClass = "hep-th",
    doi = "10.1007/JHEP08(2010)131",
    journal = "JHEP",
    volume = "08",
    pages = "131",
    year = "2010"
}

@article{Anninos:2009yc,
    author = "Anninos, Dionysios and Hartman, Thomas",
    title = "{Holography at an Extremal De Sitter Horizon}",
    eprint = "0910.4587",
    archivePrefix = "arXiv",
    primaryClass = "hep-th",
    doi = "10.1007/JHEP03(2010)096",
    journal = "JHEP",
    volume = "03",
    pages = "096",
    year = "2010"
}

@article{Christodoulou:1991cr,
    author = "Christodoulou, D.",
    title = "{Nonlinear nature of gravitation and gravitational wave experiments}",
    doi = "10.1103/PhysRevLett.67.1486",
    journal = "Phys. Rev. Lett.",
    volume = "67",
    pages = "1486--1489",
    year = "1991"
}

@article{Hintz:2026jiw,
    author = "Hintz, Peter",
    title = "{Perturbations of Kerr-Newman black holes: separation and mode stability}",
    eprint = "2609.33661",
    archivePrefix = "arXiv",
    primaryClass = "gr-qc",
    month = "9",
    year = "2026"
}

@article{Casals:2024ynr,
    author = "Casals, Marc and Hollands, Stefan and Pound, Adam and Toomani, Vahid",
    title = "{Spin-2 Green{\textquoteright}s functions on Kerr in radiation gauge}",
    eprint = "2402.15468",
    archivePrefix = "arXiv",
    primaryClass = "gr-qc",
    doi = "10.1088/1361-6382/ad7cbc",
    journal = "Class. Quant. Grav.",
    volume = "41",
    number = "21",
    pages = "215010",
    year = "2024"
}
\appendix

\newpage
\onecolumngrid
\newpage
\section{End Matter}
\subsection{Background Newman--Penrose quantities and Operators}
Varying the Einstein--Maxwell action, given in Eq.~\eqref{eq: EinsteinMaxwellAction}, yields the following equations of motion.
\begin{subequations}
\begin{align}\label{eq: EinsteinMaxwellEquOfMot}
G_{\mu \nu}+\Lambda_c g_{\mu \nu} & =2\left(F_{\mu \rho} F_\nu^\rho-\frac{1}{4} g_{\mu \nu} F_{\rho \sigma} F^{\rho \sigma}\right), \\
\nabla_\mu F^{\mu \nu} & =
0 
\end{align}
\end{subequations} 
The solutions we consider at $\mathcal{O}(\epsilon^0)$ are given by the Kerr--Newman--de~Sitter black hole line element
\begin{align}\label{eq: KNdSLineElemenet}
    d s^2=  \rho_K^2 \left(\frac{d r^2}{\Delta_r}+\frac{d \theta^2}{\Delta_\theta}\right)+\frac{\Delta_\theta \sin ^2 \theta}{(1+\alpha_c)^2 \rho_K^2}\left[a d t-\left(r^2+a^2\right) d \varphi\right]^2 -\frac{\Delta_r}{(1+\alpha_c)^2 \rho_K^2}\left(d t-a \sin ^2 \theta d \varphi\right)^2\,,
\end{align}
along with its corresponding gauge potential given by
\begin{align}\label{eq: KNdSGaugeField}
    A_\mu d x^\mu &=-\frac{Q r}{(1+\alpha_c) \rho_K^2}\left(d t-a \sin ^2 \theta d \varphi\right) \,,
\end{align}
where
\begin{align}\label{eq: KNMetricQuants}
\Delta_r=\left(r^2+a^2\right)\left(1-\alpha_c r^2 / a^2\right)-2 M r+Q^2; \quad 
\alpha_c=\Lambda_c a^2 / 3;\quad \Delta_\theta=1+\alpha_c \cos ^2 \theta; \quad 
\rho_K^2=r^2+a^2 \cos ^2 \theta 
\end{align}
We use the following definitions of the Weyl and Maxwell scalars:
\begin{subequations}
\begin{align}
\Psi_0 = C_{lmlm},
\quad
\Psi_1 = C_{lnlm},
\quad
\Psi_2 &= C_{lm\bar m n},
\quad
\Psi_3 = C_{ln\bar m n},
\quad
\Psi_4 = C_{n\bar m n\bar m},
\label{eq: WeylNPScalars}
\\
\Phi_0 = F_{lm},
\quad
\Phi_1& = \frac12\left(F_{ln}+F_{\bar m m}\right),
\quad
\Phi_2 = F_{\bar m n}.
\label{eq: MaxwellNPScalars}
\end{align}
\end{subequations}
where $C$ and $F$ are the Weyl and Maxwell field-strength tensors, respectively. Writing $\Sigma\equiv\rho_K^2$, the principal null tetrad is given by \cite{Kraniotis:2018zmh}
\begin{align}\label{eq: KNDSFrame}
l^\mu=\frac{\bigl(\Xi(r^2+a^2),\Delta_r,0,a\Xi\bigr)}{\Delta_r},
\quad n^\mu=\frac{\bigl(\Xi(r^2+a^2),-\Delta_r,0,a\Xi\bigr)}{2\Sigma},
\quad m^\mu=-\frac{\bar\rho}{\sqrt{2\Delta_\theta}}
\bigl(ia\Xi\sin\theta,0,\Delta_\theta,i\Xi\csc\theta\bigr)
\end{align}
where $\Xi \equiv 1 + \alpha_c$. In this frame, the nonvanishing background Weyl, Maxwell, and Ricci scalars are
\begin{equation}\label{eq:KNdSBackgroundScalars}
\begin{aligned}
\Psi_2
&=\rho^3\left(M+Q^2\bar\rho\right),
&\qquad
\Phi_1
&=\frac{Q}{2}\rho^2,
\\
\Phi_{11}
&=2\Phi_1\bar{\Phi}_1
=\frac{Q^2}{2}|\rho|^4,
&\qquad
\Lambda_{\mathrm{NP}}
&=\frac{R}{24}=\frac{\Lambda_c}{6}.
\end{aligned}
\end{equation}
where $\rho = -(r - ia \cos\theta)^{-1}$.

The Teukolsky operators\footnote{Notice that these operators correspond to the operators of Ref.~\cite{Dias:2015wqa} after a Geroch--Held--Penrose \cite{Geroch:1973am, held1974formalism} transformation.} of Eq.~\eqref{eq: OutgoingKNEqus} are given by
\begin{subequations}\label{eq: PosistiveSpinOperators}
\begin{align}
\mathcal O_{+2}
&=D_{[-3,1,-4,-1]}\Delta_{[1,0,-4,0]}
-\delta_{[-1,-3,1,-4]}\bar\delta_{[-4,0,1,0]}
-3\Psi_2,
\\
\mathcal P_{+2}
&=2
-4D_{[-3,1,0,0]}A_-\Delta_{[1,0,-4,0]}
-4(\tau-\bar\pi)A_+\bar\delta_{[-4,0,1,0]},
\\
\mathcal Q_{+2}
&=\frac{2}{\Phi_1}\Big[
D_{[-3,1,-2,0]}A_-\delta_{[0,-2,0,-6]}
+(\tau-\bar\pi)A_+D_{[-2,0,-6,0]}
\Big],
\\
\mathcal O_{+1}
&=D_{[-1,1,-5,-1]}\Delta_{[4,0,-2,0]}
-\delta_{[-1,-1,1,-5]}\bar\delta_{[-2,0,4,0]},
\\
\mathcal P_{+1}
&=2\Delta_{[4,-1,-3,-1]}A_+D_{[-2,0,-6,0]}
+2\bar\delta_{[-3,1,4,1]}A_-\delta_{[0,-2,0,-6]},
\\
\mathcal Q_{+1}
&=-4\Phi_1\Big[
\Delta_{[2,-1,-3,-1]}A_+\bar\delta_{[-4,0,1,0]}
+\bar\delta_{[-3,1,2,1]}A_-\Delta_{[1,0,-4,0]}
\Big].
\end{align}
\end{subequations}
where $A_{ \pm}=\left(3 \Psi_2 \mp 2 \Phi_{11}\right)^{-1} $. Maxwell's equations for Petrov type D spacetimes are given by the following
\begin{subequations}\label{eq: NPMaxwellEqusEndMatter}
\begin{align}
D_{[0,0,-2,0]}\Phi_1
-\bar{\delta}_{[-2,0,1,0]}\Phi_0
&=2\pi J_l,
&\qquad
\delta_{[0,0,0,-2]}\Phi_1
-\Delta_{[1,0,-2,0]}\Phi_0
&=2\pi J_m,
\label{eq: MaxwellOneTwoEM}\\
D_{[2,0,-1,0]}\Phi_2
-\bar{\delta}_{[0,0,2,0]}\Phi_1
&=2\pi J_{\bar m},
&\qquad
\delta_{[0,2,0,-1]}\Phi_2
-\Delta_{[2,0,0,0]}\Phi_1
&=2\pi J_n.
\label{eq: MaxwellThreeFourEM}
\end{align}
\end{subequations}
where $J_{a} = J_{\mu}e^{\mu}_a$ is the current in the Newman--Penrose basis. 

\subsection{Reconstruction Steps}

Using the frame gauge of Eq.~\eqref{eq: FrameGaugeIRG}, some of the spin coefficients that we use in our procedure can be expressed in terms of the metric perturbation in the ingoing radiation gauge as follows \cite{Wagle:2023fwl}:
\begin{subequations}\label{eq:ReconstructuedSpinCoeifEM}
\begin{alignat}{2}
\sigma^{(1)}
&= -\frac{1}{2}
D_{[-2,2,1,-1]} h_{mm}^{(1)},
&\qquad
\Psi_0^{(1)}
&= -\frac{1}{2}
D_{[-3,1,0,-1]} D_{[-2,2,0,-1]} h_{mm}^{(1)},
\\[3pt]
\tau^{(1)}
&= \frac{1}{2}\Bigl[
-D_{[0,2,0,-1]} h_{nm}^{(1)}
+\pi h_{mm}^{(1)}
\Bigr],
&\qquad
\bar{\lambda}^{(1)}
&= (\bar{\pi}+\tau)h_{nm}^{(1)}
+\frac{1}{2}
\Delta_{[1,-1,-2,2]} h_{mm}^{(1)},
\\[3pt]
\bar{\alpha}^{(1)}
&= \frac{1}{4}\Bigl[
-D_{[0,-2,-2,-1]} h_{nm}^{(1)}
+\bar{\delta}_{[-2,0,1,1]} h_{mm}^{(1)}
\Bigr],
&\qquad
\beta^{(1)}
&= -\frac{1}{4}\Bigl[
D_{[-4,2,2,-1]} h_{nm}^{(1)}
+\bar{\delta}_{[0,2,-1,-1]} h_{mm}^{(1)}
\Bigr].
\end{alignat}
\end{subequations}
The remaining Weyl scalar needed for the transport equations is
\begin{equation}\label{eq: Psi1IRG}
\begin{aligned}
\Psi_1^{(1)}
= -\frac{1}{8}\Bigl[2D_{[-1,1,1,-1]} D_{[0,2,1,-1]} h_{nm}^{(1)}
+D_{[-1,1,1,-1]} \bar{\delta}_{[-2,2,-2,-1]} h_{mm}^{(1)}
+\bar{\delta}_{[-3,1,-3,-1]}
 D_{[-2,2,0,-1]} h_{mm}^{(1)}
\Bigr].
\end{aligned}
\end{equation}

The transport equations required to obtain $\Phi_0^{(1)}$ and $\Psi_2^{(1)}$,
Eqs.~\eqref{eq: Phi0Transport} and~\eqref{eq: Psi2Transport},
follow from the linearization of the respective Bianchi identities
\cite{Chandrasekhar_1983,Li:2026rkf}
\begin{align}
\bar{\delta}_{[-4,0,1,0]}\Psi_0
-D_{[-2,0,-4,0]}\Psi_1
-3\kappa\Psi_2
&= \mathcal{R}_a,
\label{eq: Bianchihnm}
\\
D_{[0,0,-3,0]}\Psi_2
-\bar{\delta}_{[-2,0,2,0]}\Psi_1
+\lambda\Psi_0
+2\kappa\Psi_3
&= \mathcal{R}_b,
\label{eq: Bianchihnn}
\end{align}
where the source terms are given by
\begin{subequations}
\begin{align}
\mathcal{R}_a
&\equiv
\delta_{[-2,-2,1,0]}\Phi_{00}
-D_{[-2,0,0,-2]}\Phi_{01}
+2\sigma\Phi_{10}
-2\kappa\Phi_{11}
-\bar{\kappa}\Phi_{02},
\\
\mathcal{R}_b
&\equiv
-\Delta_{[0,1,-2,-2]}\Phi_{00}
+\bar{\delta}_{[-2,0,0,-2]}\Phi_{01}
+2\rho\Phi_{11}
+\bar{\sigma}\Phi_{02}
-2\tau\Phi_{10}
-2D\Lambda_{\rm NP}.
\end{align}
\end{subequations}
In the chosen frame gauge of Eq.~\eqref{eq: FrameGaugeIRG}, using
$\Phi_{00}^{(1)}=0$ and $\Lambda_{\rm NP}^{(1)}=0$, the source in
Eq.~\eqref{eq: Psi2Transport} becomes
\begin{equation}\label{eq: SourcePsi_2}
\mathcal{S}_{\Psi_2}^{(1)}
\bigl[\Psi^{(1)}_1,\Phi^{(1)}_{01},\Phi^{(1)}_{11},\Phi^{(1)}_{10}\bigr]
=
\bar{\delta}_{[-2,0,2,0]}\Psi_1^{(1)}
+\bar{\delta}_{[-2,0,0,-2]}\Phi_{01}^{(1)}
+2\rho\Phi_{11}^{(1)}
-2\tau\Phi_{10}^{(1)}.
\end{equation}

Once $\Psi_2^{(1)}$ has been determined, the remaining metric
component follows from the Ricci identity
\begin{equation}\label{eq: Riccihnn}
\Delta\rho-\bar{\delta}\tau
=
-\rho\bar{\mu}-\sigma\lambda
+(\bar{\beta}-\alpha-\bar{\tau})\tau
+(\gamma+\bar{\gamma})\rho
+\nu\kappa-\Psi_2-2\Lambda_{\rm NP}.
\end{equation}
Using $\rho^{(1)}=0$ in the chosen frame gauge and
$\Lambda_{\rm NP}^{(1)}=0$, its first-order expansion on the type-D
background is
\begin{equation}
\begin{aligned}
\Delta^{(1)}\rho
-\bar{\delta}^{(1)}\tau
-\bar{\delta}\tau^{(1)}
={}&
-\rho\bar{\mu}^{(1)}
+\bigl(
\bar{\beta}^{(1)}-\alpha^{(1)}-\bar{\tau}^{(1)}
\bigr)\tau
\\
&+(\bar{\beta}-\alpha-\bar{\tau})\tau^{(1)}
+(\gamma^{(1)}+\bar{\gamma}^{(1)})\rho
-\Psi_2^{(1)}.
\end{aligned}
\end{equation}
Substituting the metric expressions for the perturbed directional
derivatives and spin coefficients gives
$\rho D_{[0,0,1,-1]}h_{nn}^{(1)}
=\mathcal{S}_{nn}^{(1)}$, with
\begin{equation}\label{eq: hnnSourceTerm}
\begin{aligned}
\mathcal{S}_{nn}^{(1)}
\bigl[\Psi_2^{(1)},h_{nm}^{(1)},h_{mm}^{(1)}\bigr]
={}&
-2\Psi_2^{(1)}
-\bar{\delta}_{[-1,1,0,-1]}
 D_{[0,2,0,-1]}h_{nm}^{(1)}
\\
&+\bar{\delta}_{[-1,1,0,-1]}
 \bigl(\pi h_{mm}^{(1)}\bigr)
-\rho\bar{\delta}_{[0,2,-2,-1]}h_{nm}^{(1)}
\\
&+\left(\tau D+\rho\delta_{[0,2,0,2]}\right)
 h_{n\bar m}^{(1)}
-\delta_{[0,2,0,0]}
 \left(\tau h_{\bar m\bar m}^{(1)}\right).
\end{aligned}
\end{equation}
The remaining electromagnetic relation connects the Maxwell scalar
$\Phi_1$ to the tetrad components of the gauge potential. Writing the Maxwell scalar $\Phi_1$ in terms of gauge potential components yields \cite{Wardell:2020naz}

\begin{align}
    \Phi_1
=\frac{1}{2}\Bigl[
D_{[1,1,1,-1]}A_n
-\Delta_{[-1,1,-1,-1]}A_l
+\bar{\delta}_{[-1,1,-1,-1]}A_m
-\delta_{[-1,1,1,1]}A_{\bar m}
\Bigr].
\end{align}

\newpage
\section{Supplemental Material}
\subsection{Derivation of the coupled master equations}
\label{sec: derivationSec}
Here, we detail the derivation of the master equations, given in Eq.~\eqref{eq: OutgoingKNEqus} of the main text. We demonstrate these equations are also valid for linear perturbations of Kerr--Newman--de Sitter where previously they were shown to be valid for the Kerr--Newman background \cite{lee1976coupled, Dias:2015wqa}. 

Taking the Kerr--Newman--de Sitter black hole as our background solution at $\mathcal{O}(\epsilon^0)$ we recall the following are satisfied $\Lambda_{\rm NP}=\Lambda_c/6$,
 and $\nabla_\mu\Lambda_{\rm NP}=0$ while at leading order the equations of motion impose $\Lambda_{\rm NP}^{(1)}=0$.
 The two Bianchi identities before linearization are \cite{Li:2022pcy}
\begin{subequations}\label{eq:KNdSFullBianchi}
\begin{align}
\bar\delta_{[-4,0,1,0]}\Psi_0
-D_{[-2,0,-4,0]}\Psi_1
-3\kappa\Psi_2
&=\mathcal{R}_\kappa\,,
\\
\Delta_{[1,0,-4,0]}\Psi_0
-\delta_{[0,-2,0,-4]}\Psi_1
-3\sigma\Psi_2
&=\mathcal{R}_\sigma\,,
\end{align}
\end{subequations}
where the source terms are given by
\begin{subequations}\label{eq:KNdSFullBianchiSources}
\begin{align}
\mathcal{R}_\kappa
&\equiv
\delta_{[-2,-2,1,0]}\Phi_{00}
-D_{[-2,0,0,-2]}\Phi_{01}
+2\sigma\Phi_{10}
-2\kappa\Phi_{11}
-\bar\kappa\Phi_{02}\,,
\\
\mathcal{R}_\sigma
&\equiv
\delta_{[0,-2,2,0]}\Phi_{01}
-D_{[-2,2,0,-1]}\Phi_{02}
-\bar\lambda\Phi_{00}
+2\sigma\Phi_{11}
-2\kappa\Phi_{12}\,,
\end{align}
\end{subequations}
which does not have any terms with $\Lambda_{\rm NP}$.

In the notation of the main text, set
$f=\Phi_0^{(1)}/(2\Phi_1)$ and
$A_\pm=(3\Psi_2\mp2\Phi_{11})^{-1}$.
Expanding up to $\mathcal{O}(\epsilon^1)$, we can express each identity in terms of the master variables $\varphi^{(1)}_{+2}$ and $\varphi^{(1)}_{+1}$ and 
\begin{subequations}\label{eq:KNdSBianchiCombinations}
\begin{align}
\mathcal L_\kappa
&\equiv A_+\left[
 \bar\delta_{[-4,0,1,0]}\varphi_{+2}^{(1)}
 -D_{[-2,0,-4,0]}\left(\frac{\varphi_{+1}^{(1)}}{2\Phi_1}\right)
 \right]
 =\kappa^{(1)}+D_{[-2,0,-1,0]}f,\\
\mathcal L_\sigma
&\equiv A_-\left[
 \Delta_{[1,0,-4,0]}\varphi_{+2}^{(1)}
 -\delta_{[0,-2,0,-4]}\left(\frac{\varphi_{+1}^{(1)}}{2\Phi_1}\right)
 \right]
 =\sigma^{(1)}+\delta_{[0,-2,0,-1]}f\,,
\end{align}
\end{subequations}

Apply $D_{[-3,1,-1,-1]}$ to $\mathcal L_\sigma$ and subtract
$\delta_{[-1,-3,1,-1]}$ applied to the $\mathcal L_\kappa$ identity.
The auxiliary field $f$ cancels by the type-D commutator identity
\cite{Teukolsky:1973ha}
\begin{equation}
D_{[-(p+1),1,q,-1]}\delta_{[0,-p,0,q]}
-\delta_{[-1,-(p+1),1,q]}D_{[-p,0,q,0]}=0,
\label{eq:SMmasterCommutator}
\end{equation}
with $(p,q)=(2,-1)$. The remaining spin-coefficient terms satisfy
the linearized optical Ricci identity
\begin{equation}
D_{[-3,1,-1,-1]}\sigma^{(1)}
-\delta_{[-1,-3,1,-1]}\kappa^{(1)}
=\varphi_{+2}^{(1)}.
\label{eq:SMmasterOptical}
\end{equation}
Therefore, the first master equation is given by
\begin{equation}
\mathscr E_{+2}\equiv
D_{[-3,1,-1,-1]}\mathcal L_\sigma
-\delta_{[-1,-3,1,-1]}\mathcal L_\kappa
-\varphi_{+2}^{(1)}=0.
\label{eq:SMmasterSpin2Factorized}
\end{equation}

Eq.~\eqref{eq: OutgoingKNEqus} of the main text, we need to consider the following Maxwell relations
\begin{subequations}\label{eq:SMmasterFullMaxwell}
\begin{align}
D_{[0,0,-2,0]}\Phi_1
&=\bar\delta_{[-2,0,1,0]}\Phi_0-\kappa\Phi_2,
\\
\delta_{[0,0,0,-2]}\Phi_1
&=\Delta_{[1,0,-2,0]}\Phi_0-\sigma\Phi_2.
\end{align}
\end{subequations}

Apply $\delta_{[-1,-1,1,-2]}$ to the first equation and
$D_{[-1,1,-2,-1]}$ to the second, and subtract. Both terms
$\kappa\Phi_2$ and $\sigma\Phi_2$ are quadratic in the perturbations.
Define the operator
\begin{equation}
\mathcal M_0\equiv
\delta_{[-1,-1,1,-2]}\bar\delta_{[-2,0,1,0]}
-D_{[-1,1,-2,-1]}\Delta_{[1,0,-2,0]}.
\label{eq:SMmasterM0}
\end{equation}
We use the following relations
\begin{subequations}\label{eq:SMMaxwellGeometricIdentities}
\begin{align}
[\delta,D]
&=(\bar\alpha+\beta-\bar\pi)D+\kappa\Delta-(\bar\rho+\varepsilon-\bar\varepsilon)\delta
-\sigma\bar\delta,
\label{eq:SMMaxwellDeltaD}
\\
D\tau-\Delta\kappa
&=(\tau+\bar\pi)\rho+(\bar\tau+\pi)\sigma
+(\varepsilon-\bar\varepsilon)\tau-(3\gamma+\bar\gamma)\kappa+\Psi_1+\Phi_{01},
\label{eq:SMMaxwellRicciTau}
\\
\delta\rho-\bar\delta\sigma
&=(\bar\alpha+\beta)\rho-(3\alpha-\bar\beta)\sigma
+(\rho-\bar\rho)\tau+(\mu-\bar\mu)\kappa-\Psi_1+\Phi_{01}.
\label{eq:SMMaxwellRicciRho}
\end{align}
\end{subequations}
Subtracting Eq.~\eqref{eq:SMMaxwellRicciRho} from
Eq.~\eqref{eq:SMMaxwellRicciTau} and rearranging gives
\begin{equation}\label{eq:SMMaxwellRicciDifference}
\begin{aligned}
&D\tau-\delta\rho
+(\bar\alpha+\beta-\bar\pi)\rho
-(\varepsilon-\bar\varepsilon+\bar\rho)\tau=
\Delta_{[-1,1,-3,-1]}\kappa
-\bar\delta_{[-3,1,-1,-1]}\sigma
+2\Psi_1.
\end{aligned}
\end{equation}

When applying Eq.~\eqref{eq:SMMaxwellDeltaD} to $\Phi_1$ before
linearization, the terms $\kappa\Delta-\sigma\bar\delta$ must be
retained. Although $\kappa=\sigma=0$ on the aligned background,
they contribute
$\kappa^{(1)}\Delta\Phi_1-\sigma^{(1)}\bar\delta\Phi_1$
at first order, where the unmarked quantities in this last expression
are evaluated on the background.
Expanding we obtain
\begin{equation}
\begin{aligned}
\mathcal M_0\Phi_0^{(1)}
={}&\kappa^{(1)}\Delta\Phi_1
-\sigma^{(1)}\bar\delta\Phi_1+2\Phi_1\left[
\Delta_{[-1,1,-3,-1]}\kappa^{(1)}
-\bar\delta_{[-3,1,-1,-1]}\sigma^{(1)}
+2\Psi_1^{(1)}
\right].
\end{aligned}
\label{eq:SMmasterMaxwellIntermediate}
\end{equation}
Using the background Maxwell relations, where $\Delta\Phi_1=-2\mu\Phi_1$ and
$\bar\delta\Phi_1=-2\pi\Phi_1$, then leads to
\begin{equation}
\frac{\mathcal M_0\Phi_0^{(1)}}{2\Phi_1}
=\Delta_{[-2,1,-3,-1]}\kappa^{(1)}
-\bar\delta_{[-3,1,-2,-1]}\sigma^{(1)}
+2\Psi_1^{(1)}.
\label{eq:SMmasterMaxwellClosure}
\end{equation}

Expressing this in terms of the master variable $\varphi^{(1)}_{+1}$ and using Eq.~\eqref{eq:KNdSBianchiCombinations} we obtain
\begin{equation}
\mathscr E_{+1}\equiv
\Delta_{[-2,1,-3,-1]}\mathcal L_\kappa
-\bar\delta_{[-3,1,-2,-1]}\mathcal L_\sigma
+\frac{\varphi_{+1}^{(1)}}{\Phi_1}=0\,,
\label{eq:SMmasterSpin1Factorized}
\end{equation}
yielding the second master equation.

\subsection{Parity Conjugate Relations}\label{sec: ParitySection}

The twelve background spin coefficients are
\begin{equation}\label{eq:KNdSBackgroundSpinCoefficients}
\begin{aligned}
&\kappa=\sigma=\lambda=\nu=\varepsilon=0,
\\[2pt]
&\rho=-\frac{1}{r-ia\cos\theta},
\qquad
\mu=\frac{\Delta_r}{2}\rho|\rho|^2,
\qquad
\gamma=\mu+\frac{\Delta_r'}{4}|\rho|^2,
\\[2pt]
&\pi=\frac{ia\sin\theta\sqrt{\Delta_\theta}}{\sqrt{2}}\,\rho^2,
\qquad
\tau=-\frac{ia\sin\theta\sqrt{\Delta_\theta}}{\sqrt{2}}\,|\rho|^2,
\\[2pt]
&\beta=
-\frac{\bar\rho}{4\sqrt{2\Delta_\theta}}
\left(
2\Delta_\theta\cot\theta+\partial_\theta\Delta_\theta
\right),
\\[2pt]
&\alpha=
\pi+\frac{\rho}{4\sqrt{2\Delta_\theta}}
\left(
2\Delta_\theta\cot\theta+\partial_\theta\Delta_\theta
\right)
=\pi-\bar\beta \,
\end{aligned}
\end{equation}
where $\Delta_r'=d\Delta_r/dr$ and
$|\rho|^2=\rho\bar\rho=(r^2+a^2\cos^2\theta)^{-1}$.

Following Ref.~\cite{Li:2023ulk} and the definitions given in the main text, define parity conjugation by $\hat{\mathcal P}f(t,r,\theta,\phi)
=
\overline{f(t,r,\pi-\theta,\phi+\pi)}$, where $\hat{P}\hat{C} = \hat{\mathcal{P}}$ and $\hat{P}$ and $\hat{C}$ are the ordinary parity and complex-conjugate operator. For the tetrad given in Eq.~\eqref{eq: KNDSFrame}, satisfies
$\hat{P}\ell^{\mu}=\ell^{\mu}$, $\hat{P}n^{\mu}=n^{\mu}$, and $\hat{P}m^{\mu}=-\bar m^{\mu}$.
Consequently, the background spin coefficients obey
\begin{equation}\label{eq:PCSpinCoefficients}
\begin{aligned}
\hat{\mathcal P}\{\sigma,\lambda,\varepsilon,\rho,\mu,\gamma\}
&=
+\{\sigma,\lambda,\varepsilon,\rho,\mu,\gamma\},
\\
\hat{\mathcal P}\{\kappa,\nu,\alpha,\beta,\tau,\pi\}
&=
-\{\kappa,\nu,\alpha,\beta,\tau,\pi\}.
\end{aligned}
\end{equation}
The same signs apply to their complex conjugates.
Each angular tetrad leg, including a directional-derivative leg,
contributes a minus sign; the entries for vanishing coefficients
record this geometric transformation rule.
The modified derivatives defined in the main text satisfy, for real
$a,b,c,d$,
\begin{subequations}\label{eq:PCModifiedDerivatives}
\begin{align}
\hat{\mathcal P}\!\left[D_{[a,b,c,d]}\right]
&=D_{[a,b,c,d]},
&
\hat{\mathcal P}\!\left[\Delta_{[a,b,c,d]}\right]
&=\Delta_{[a,b,c,d]},
\\
\hat{\mathcal P}\!\left[\delta_{[a,b,c,d]}\right]
&=-\delta_{[a,b,c,d]},
&
\hat{\mathcal P}\!\left[\bar\delta_{[a,b,c,d]}\right]
&=-\bar\delta_{[a,b,c,d]}.
\end{align}
\end{subequations}
At $\mathcal{O}(\epsilon^1)$ we also find in the ingoing radiation gauge
\begin{align}\label{eq:PCModifiedDerivativesFO}
\hat{\mathcal P}\left[\Delta  ^{(1)}_{[a,b,c,d]}\right]
=\pm (-1)^{\ell}\left[\Delta  ^{(1)}_{[a,b,c,d]}\right] \quad;\quad\hat{\mathcal P}\left[\bar{\delta} ^{(1)}_{[a,b,c,d]}\right]
=\mp (-1)^{\ell}\left[\bar{\delta}  ^{(1)}_{[a,b,c,d]}\right] \quad;\quad\hat{\mathcal P}\left[\delta  ^{(1)}_{[a,b,c,d]}\right]
=\mp (-1)^{\ell}\left[\delta^{(1)}_{[a,b,c,d]}\right] 
\end{align}
where we have not included $D^{(1)}_{[a,b,c,d]}$ since it vanishes in this gauge.

\end{document}